\makeatletter
\pdfoutput=1

\def\cl@chapter{}
\makeatother
\documentclass[epj]{svjour}

\usepackage{xcolor}

\usepackage{url}
\usepackage[utf8]{inputenc}
\usepackage{graphicx}
\usepackage{braket}
\usepackage{xintexpr}
\usepackage{siunitx}
\usepackage{amsmath,amssymb,amsfonts}%
\usepackage{xspace}
\usepackage{hyperref}
\usepackage{cleveref}

\usepackage{multirow}%
\usepackage{mathrsfs}%
\usepackage[title]{appendix}%
\usepackage{xcolor}%
\usepackage{textcomp}%
\usepackage{manyfoot}%
\usepackage{booktabs}%
\usepackage{algorithm}%
\usepackage{algorithmicx}%
\usepackage{algpseudocode}%
\usepackage{listings}%

\newcolumntype{C}{>{$}c<{$}}

\begin{document}

\title{Finding excesses in model parameter space 
}

\author{Kierthika Chathirathas\inst{1}\thanks{kierthika.chathirathas@kit.edu}, Torben~Ferber\inst{2}\thanks{torben.ferber@kit.edu}, Felix~Kahlhoefer\inst{3}\thanks{kahlhoefer@kit.edu} and Alessandro~Morandini \inst{1}\thanks{alessandro.morandini@kit.edu}}

\institute{Institute for Astroparticle Physics (IAP),  Karlsruhe Institute of Technology (KIT), D-76131 Karlsruhe, Germany \and Institute of Experimental Particle Physics (ETP),  Karlsruhe Institute of Technology (KIT), D-76131 Karlsruhe, Germany \and Institute for Theoretical Particle Physics (TTP),  Karlsruhe Institute of Technology (KIT), D-76131 Karlsruhe, Germany}

\authorrunning{K. Chathirathas, T. Ferber, F. Kahlhoefer, A. Morandini}
\titlerunning{Finding excesses in model parameter space}

\abstract{Simulation-based inference (SBI) makes it possible to infer the parameters of a model from high-dimensional low-level features of the observed events. In this work we show how this method can be used to establish the presence of a weak signal on top of an unknown background, to discard background events and to determine the signal properties. The key idea is to use SBI methods to identify events that are similar to each other in the sense that they agree on the inferred model parameters. We illustrate this method for the case of axion-like particles decaying to photons at beam-dump experiments. For poor detector resolution the diphoton mass cannot be reliably reconstructed, so there is no simple high-level observable that can be used to perform a bump hunt. Since the SBI methods do not require explicit high-level observables, they offer a promising alternative to increase the sensitivity to new physics.}


\maketitle

\section{Introduction}

A complete experimental analysis turns observations into either exclusion limits on, or preferred regions in the parameter space of a model, depending on whether or not there is a (significant) excess over expected backgrounds. A well-established and powerful procedure to perform this kind of analysis is bump hunting~\cite{Choudalakis:2011qn}, which searches for a signal that is localised (in terms of some informative observables) compared to a more broadly distributed background. The central advantage of bump hunting is that it can be performed even in situations where the background cannot be reliably simulated or contains a-priori unknown contributions, as long as it is sufficiently smooth. If an excess over the background is observed, it provides two pieces of information. First, it may be an indication that the background-only hypothesis is not the right explanation for the data. And second, the position of the excess is informative about the properties of a hypothetical signal. 

While bump hunting is a very powerful technique and has lead to the discovery of various new particles, it suffers from a number of limitations that need to be addressed. One of them is the look-elsewhere effect \cite{Gross:2010qma} related to the unknown location of the hypothetical signal ahead of the experiment. Moreover, bump hunting works best if the number of signal events is large, in the sense that the separation between two neighbouring events (in the observable under consideration) is small compared to the experimental resolution. The opposite case, where individual events are well-separated, makes it challenging to obtain numerically stable inferences for the background and signal model.

For the success of bump hunting it is crucial to identify a suitable observable and construct a summary statistic. For example, when searching for a new particle decaying into two final state particles that can be well-measured in the detector, the obvious choice is the invariant mass of the decay products. However, in more complicated scenarios, it can be difficult to find the optimal way to combine low-level features into one or more high-level observables. As shown in Ref.~\cite{Morandini:2023pwj}, even for a two-body decay the invariant mass may not be the optimal choice, if the four-vectors of the final state particles are hard to measure accurately.

In such a scenario, the optimal sensitivity to new physics may be obtained by directly analysing the low-level features of individual events using the methods of simulation-based inference (SBI), i.e.\ machine-learning (ML) algorithms  that do not explicitly construct high-level observables nor summary statistics, which are conventionally needed to calculate likelihoods \cite{cranmer2020frontier,Brehmer:2020cvb,hermans2021trust,Papamakarios:2016ctj,Rizvi:2023mws}. Instead, SBI algorithms need simulations where the observed data are paired with the model parameters used to generate the simulations. Once trained, these methods can then be applied to a data set to directly provide constraints on the parameter space of a model in terms of likelihood ratios for the frequentist case or likelihood-to-evidence ratios for the Bayesian case. However, due to their reliance on simulations, it is immediately not obvious how to apply SBI methods to the case of unknown or difficult-to-simulate backgrounds.

In this work we propose two new ML methods that combine the advantages of SBI and bump hunting. The central idea is to search for events that are compatible regarding their low-level features, in the sense that these events would allow for a common interpretation in terms of a specific signal hypothesis. Intuitively, instead of performing a bump hunt in a high-level observable, we perform a bump hunt in model parameter space. The signal is in this case identified not by its qualitative difference from the background, but rather by its rarity \cite{Caron:2021wmq}: Under the background-only hypothesis there is a low probability of having multiple background events compatible with the same set of model parameters.

There has been extensive work in particle physics to address the identification of a signal on top of the background and we refer the reader to Refs.~\cite{Golling:2023yjq,Belis:2023mqs} for comprehensive reviews. However, most of the methods in these studies rely on the presence of sizable side-bands to extend the study to the signal region \cite{Golling:2022nkl,Andreassen:2020nkr,Hallin:2021wme,Hallin:2022eoq,Raine:2022hht}. Another type of approach, more similar to ours in the underlying idea, is called New Physics Learning Machine (NPLM) and uses the test statistic created by a network to establish the compatibility between a data sample and a reference (background) sample  \cite{dAgnolo:2021aun,DAgnolo:2018cun,DeSimone:2018efk}. However, in our approach we study the compatibility between events based on whether they can be explained by the same physical model rather than establishing the compatibility of data sets based on an ML-engineered test statistic.

Our two methods differ from each other in whether or not they explicitly perform a mapping to the parameter space of the signal hypothesis. The first method, which we call Event Compatibility based on Observables (ECO), requires no such mapping and directly works with low-level observables, whereas the second, which we call Events with Posterior Overlap (EPO) explicitly calculates posteriors for individual events. We will show that both approaches give similar results and are particularly well suited for the case of small event numbers. In the case of a detector with angular resolution worse than $\SI{10}{\milli\radian}$ they outperform the conventional analysis technique that we use for comparison and offer an exciting opportunity to improve sensitivity in future searches.

A central advantage of the ML approach is that -- once is has been trained on the whole model parameter space -- it does not require specific hypothesis regarding the location of the signal when evaluating the compatibility of different events. Our approach therefore does not suffer from a look-elsewhere effect in the same way as conventional bump hunting based on signal templates.
It is furthermore easily possible to adjust the threshold for event compatibility to account for different expected background rates and to adapt to different experimental designs. Modifying the threshold to maximise the expected sensitivity does not require any retraining of the ML models. This makes our approach a valuable tool also for sensitivity studies and experimental design.

As a physically motivated application we consider a search for axion-like particles (ALPs) decaying into photons in proton beam-dump experiments, which are particularly sensitive to feeble interactions thanks to their high beam intensity. Our focus will be on the SHiP experiment~\cite{SHiP:2021nfo}, which has been selected from several proposals~\cite{Ahdida:2023okr} as the successor of the NA62 experiment at CERN~\cite{NA62:2017rwk}. SHiP is designed to be a zero-background experiment, such that a very small number of observed events could potentially lead to a convincing discovery if the background hypothesis can be robustly rejected. This makes ALP searches at SHiP an ideal application of our approach. However, the same ML methods can be applied to other scenarios of new physics and other types of experiments where both background and signal are small but hard to distinguish.

Our work is structured as follows. In \cref{sec:simulator} we describe our model simulations and the details of the experiment, followed in \cref{sec:methodology} by the ML algorithms used and the steps of the analysis strategy. In \cref{sec:sig_detection} we discuss how to reject the background-only hypothesis and in \cref{sec:bkg_remove} how to remove the possible background events. In \cref{sec:inference} we take advantage of the already extracted single-event posteriors to derive the model parameters. In particular, we study how the combination of the cleaned event samples gives a reliable and precise estimation of the model parameters. We finally conclude in \cref{sec:conclusions}.

\section{ALPs at beamdumps}\label{sec:simulator}
\subsection{Event generation} 

The process under investigation is the production of ALPs at proton beam-dump experiments and their subsequent decay into a pair of photons. Regarding the production mode, we focus on ALPs produced in rare meson decays, which means mainly $B$ meson decays at the energies and masses relevant for us. This scenario is realised for instance in the $W$ boson dominance scenario~\cite{Izaguirre:2016dfi}, where the ALP (denoted by $a$) couples primarily to $SU(2)_L$ gauge bosons. However, we do not assume a one-to-one correspondence between the production cross section and the decay rate and instead treat the two processes as independent. This corresponds for example to the case where an additional coupling of the ALP to hypercharge gauge bosons modifies the effective coupling to photons~\cite{Ertas:2020xcc}. Our simulations closely follow the public code ALPINIST~\cite{Jerhot:2022chi}, to which we refer readers interested in more technical details or different ALP models.

As input for our signal simulations, we provide the energy distribution of ALPs produced by the process $B\rightarrow Ka$ and $B \rightarrow K^\ast a$. The long-lived ALP then propagates through the absorber and decays into two photons inside the decay volume of the experiment. For a valid event, both photons need to reach the calorimeter, where they start showering. 
In our simulation we do not simulate the full shower and its reconstruction, but only extract the truth-level four-momentum and hit positions of the photons. We will refer to these as low-level observables in the following. The limited resolution and reconstruction efficiency of the calorimeter are included by applying smearing to our truth-level quantities (see below). For an event to be accepted, it furthermore needs to satisfy the experimental selection discussed in the following subsection. 

In order to generate an event, we need to state the ALP mass $m_a$ and lifetime $\tau_a$. The decay length distribution of a particle with momentum $\vec{p}_a$ is given by 
\begin{equation}\label{eq:decay_exp}
    p(d) = \exp\left( - \frac{d \, m_a}{p_a c \tau_a}\right) \; .
\end{equation}
Since the momentum $p_a$ depends only weakly on the ALP mass, the parameter combination that primarily governs the ALP decay length is the ratio of ALP lifetime over ALP mass. For this reason, we will take $m_a$ and $c\tau_a/m_a$ as the underlying parameters for our simulations. To generate signal events we need to specify the range and sampling of these parameters, reflecting the sensitivity of the experiment under study. For the ALP mass we choose the range $[\SI{0.1}{\giga\electronvolt}, \SI{4.5}{\giga\electronvolt}]$ and for the lifetime over mass we choose $[\SI{0.05}{\meter\per\giga\electronvolt}, \SI{500}{\meter\per\giga\electronvolt}]$. In both cases we sample these ranges with a logarithmic prior. As discussed in more detail in Ref.~\cite{Morandini:2023pwj} the choice of sampling for our simulations also determines the priors needed for a Bayesian analysis.

In our analysis we are interested in the case where the background expectation is very low, as should be the case for beam-dump experiments. However, we have to consider the possibility that a small number of background events pass all selections. In the case of vacuum in the decay volume, background events can be produced by muon and neutrino inelastic scatterings close to the decay surface, in addition to so-far unexplored processes. Further background events may be produced if the decay volume is filled with low-pressure air or helium. Given the rareness of these events, it is very difficult to produce sufficient Monte-Carlo samples or reliable templates. 

In the present work we therefore do not attempt to simulate actual background events. Instead, we assume that any background event that passes all quality requirements and signal selection cuts looks indistinguishable from an actual ALP decay. The only difference between signal and background is therefore that background events look like the decays of ALPs with \emph{randomly varying} mass and lifetime, whereas signal events look like the decays of ALPs with \emph{fixed} mass and lifetime. In other words, we simulate background events in the same way as the ALP signal discussed above, except that when we generate a set of background events we randomly choose different ALP parameters for each event. For simplicity, we assume a logarithmic distribution, i.e.\ we sample the ALP parameters logarithmically from the ranges defined above. We emphasise that this distribution could be easily modified if a reliable estimate of the background distribution becomes available.

One could also consider an alternative search strategy with more permissive selection cuts and correspondingly larger background rates. In this case some of the actually observed events may look different from any simulated events used for training, such that the output of the ML algorithm is difficult to predict. However, we focus here on the case of an experiment that aims to be background-free and therefore imposes strict requirements on the event selection, such that all observed events resemble events seen during training.

It should be clear from the discussion above that, lacking a background template, we cannot distinguish signal from background on an event-by-event basis. This situation changes as soon as we have at least two events. If both are signal events, the corresponding ALP parameters should be consistent with each other. In the case of background events, on the other hand, the parameters are expected to be different except for random coincidences. The decisive question is therefore how to determine whether two events are compatible or not. This decision needs to depend both on the details of the observed events and on the properties of the experiment under consideration.

\subsection{Detector geometry and experimental setup}

\begin{figure*}[t!]
\centering
\includegraphics[width=0.96\textwidth]{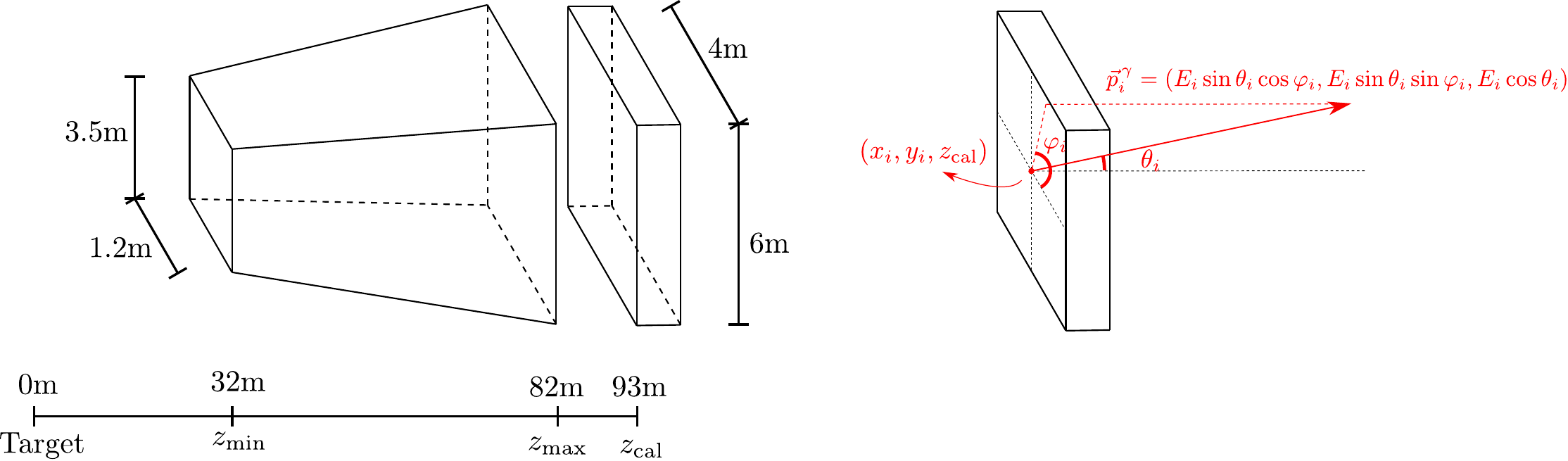}
\caption{Left: Sketch of the detector design. The calorimeter has a length in $z$ of around \SI{1}{\meter}. The system origin coincides with the proton incidence point on the target.  Right: Measurement of photon $i$ with energy $E_i$ and angles $\theta_i,\,\varphi_i$ in the calorimeter cell in position $x_i,y_i$.}
\label{fig:detectorsketch}
\end{figure*}

We focus our study on the SHiP experiment and consider the geometry examined in past sensitivity studies \cite{Ovchynnikov:2023cry}, see~\cref{fig:detectorsketch}. The experiment will have a decay volume in the shape of a truncated pyramid with two rectangular surfaces. The front surface (closer to the ALP production point) is located at $32\,\mathrm{m}$ from the target and has a size of $1.2\,\mathrm{m}\times3.5\,\mathrm{m}$. The back surface is located $82\,\mathrm{m}$ from the interaction point, such that the decay volume has a length of $50\,\mathrm{m}$. The back surface has a size of $3.9\,\mathrm{m}\times6\mathrm{m}$. Finally, at a distance of $93\,\mathrm{m}$ there is the calorimeter with a surface area of $4\,\mathrm{m}\times6\,\mathrm{m}$ and a thickness of roughly 1m. Between the decay volume and the calorimeter there are tracking chambers, which play no role for the case of photon final states. In principle the tracking chambers could be included in the decay volume, but doing so might increase the background rate. We will therefore not consider the tracking chambers as part of the decay volume.

Having simulated the ALP production and decay, we select events with a decay vertex inside the decay volume. For an event to be accepted, both photons produced in the decay must hit the calorimeter. Moreover, they need to have an energy higher than $\SI{1}{\giga\electronvolt}$ and a spatial separation larger than $d_\mathrm{min}=\SI{10}{\centi\meter}$ in the calorimeter plane. Note that we do not simulate the electromagnetic shower explicitly, but assume for simplicity that it originates on the front surface of the calorimeter. The thickness of the calorimeter is therefore irrelevant for our analysis, except that it may indirectly affect the detector resolution.

As illustrated in \cref{fig:detectorsketch}, the low-level observables obtained from our simulations are
\begin{itemize}
    \item the energies $E_1$ and $E_2$ of the two photons,
    \item their calorimeter hit positions ($x_1$, $y_1$) and $(x_2, y_2)$,
    \item the polar and azimuthal incidence angles ($\theta_1$, $\varphi_1$) and $(\theta_2$, $\varphi_2$), 
\end{itemize}
corresponding to a total of 10 features per event. 
 We find that the energy of the leading photon is quite insensitive to the ALP mass, but the opening angle of the two photons depends strongly on the ALP mass. Heavier ALPs have on average smaller boost factors and hence produce photons with larger opening angles compared to lighter ALPs. These differences however become much less pronounced once we take into account the finite detector resolution.

In addition to the geometry, the experimental setup is characterised by the calorimeter resolution. While the former is largely determined by the space available in the experimental hall and is therefore not expected to change significantly, the latter is affected by many different aspects of the calorimeter design.  We therefore consider two different benchmark scenarios for the achievable energy and angular resolution \cite{Bonivento:2018eqn}. In the good resolution scenario, we assume a relative energy uncertainty $\sigma(E)/E=0.05$ and an angular uncertainty $\sigma(\theta)=\sigma(\varphi)=\SI{5}{\milli\radian}$. In the poor resolution scenario we have $\sigma(E)/E=0.1$ and an angular uncertainty $\sigma(\theta)=\sigma(\varphi)=\SI{10}{\milli\radian}$. While the magnitude of these uncertainties is expected to be realistic, a more accurate treatment would include a dependence of both energy and angular resolution on the photon momentum. Given detailed simulations of a specific calorimeter design, such a dependence could be easily included in our analysis.
For both scenarios, we consider a fixed calorimeter hit position resolution of $\SI{1}{\milli\meter}$, since previous studies have shown that the precise hit resolution does not noticeably impact the final results~\cite{Morandini:2023pwj}.

\section{Methodology}
\label{sec:methodology}

\subsection{General considerations}

Identifying the presence of a signal on top of the background is rendered more complicated by our ignorance of where the signal should be. Even assuming a specific type of new particle, for instance an ALP, we do not know its properties. However, if a new particle is produced on-shell, its decay should result in a localised ``bump'' in some resonance feature, which can be distinguished from background. This is the procedure that lead to the Higgs discovery and the determination of the Higgs boson mass in the di-photon channel ~\cite{ATLAS:2012ae,ATLAS:2012yve,CMS:2012zhx,CMS:2013btf,ATLAS:2015yey}.

The experimental setup that we consider is rather different from the Higgs discovery, because of the much lower background level and the limitations on the experimental resolutions imposed by cost-efficiency. We are interested in the case that backgrounds are tiny, but their absence is not completely guaranteed and they are very challenging to model. This scenario includes in particular searches for long-lived particles, both at beam-dump experiments and at the LHC.

Given these differences, it is unclear whether the conventional bump hunt in the diphoton invariant mass is the optimal analysis procedure for the latter case.
We are therefore interested in alternative analysis strategies for searches for rare events with final states that are difficult to reconstruct. Our discussion will focus quantitatively on beam-dump experiments, but the discussion more generally applies to scenarios with low number of events and no background fit, for example searches for long-lived particles at the LHC.

\subsection{Conventional bump hunt}

Before turning to the new methods introduced in this work, let us briefly clarify how we perform the conventional bump hunt, which will serve as the point of comparison. As the high-level observable, we consider the diphoton invariant mass $m_{\gamma\gamma}$ reconstructed directly from the feature vectors. Given the experimental uncertainties, the measured value of $m_{\gamma\gamma}$ can be quite different from the simulated ALP mass. For example, for the case of good resolution and an ALP mass of $m_a = 1 \, \mathrm{GeV}$, the reconstructed $m_{\gamma\gamma}$ has a standard deviation of $0.25 \, \mathrm{GeV}$.

For this reason we consider the range of diphoton masses $\log_{10} m_{\gamma\gamma} / \mathrm{GeV} \in [-1.5, 0.8]$, somewhat larger than the allowed range of ALP masses. We then divide this interval into $N_\text{bins}$ subintervals and construct $N_\text{bins}$ overlapping bins by merging three adjacent subintervals. For example, for the case of $N_\text{bins} = 9$, the first bin would be $[-1.5, -0.75]$, the second bin would be $[-1.25, -0.5]$, etc. We have found that this construction performs better (and is less sensitive to the precise number of bins) than taking non-overlapping bins. The fact that bin counts in neighbouring bins are correlated will not be an issue, since we will only consider the single bin with the highest count in our analysis.

Most applications in high-energy physics would avoid binning altogether by performing an unbinned profile likelihood analysis, i.e.\ by fitting signal and background templates to the individual observed events. However, for the case of a non-Gaussian detector response, these templates are non-trivial to obtain and the fitting procedure may be numerically difficult. For simplicity, we therefore restrict ourselves to a binned analysis strategy that does not rely on templates . Given a sensible choice of binning, the two approaches are expected to give similar results.

We choose the number of bins depending on the ALP mass and detector resolution in such a way that the expected sensitivity is maximized. We use only 4 bins for $m_a=\SI{0.2}{\giga\electronvolt}$ and 11 bins for $m_a=\SI{1.0}{\giga\electronvolt}$ in both uncertainties scenarios. In the case $m_a=\SI{4.0}{\giga\electronvolt}$ we use 48 bins in the good resolution case and 30 bins in the poor resolution case.

\subsection{Machine learning models}

In this section we introduce the ML algorithms used in this work. To streamline the presentation, we focus exclusively on classifiers, which turn out to be suitable for all the tasks that we consider, noting that some of the tasks could also be performed using different architectures.  

As stressed in section~\ref{sec:simulator}, any background events passing the experimental selection are expected to look like proper signal events, except that they may not allow for a consistent interpretation in terms of an ALP with a fixed mass and lifetime. 
We will identify the events inconsistent with a common set of model parameters as background events. The way we address the compatibility is similar to contrastive learning in latent space \cite{DBLP:journals/corr/abs-1807-03748,DBLP:journals/corr/abs-1906-00910,DBLP:journals/corr/DonahueJVHZTD13,1640964,Hallin:2024gmt}. The main difference is that our target space is the model parameter space and not the latent space of a network, since we want to take advantage of our physical knowledge. The compatibility scores are then processed through an algorithm to select the ones above a certain threshold. This is analogous to what is done with clustering algorithms \cite{10.1145/1772690.1772862,Ester1996ADA,10.1145/3068335}, with our scores determining whether the events belong to the most populated cluster.

The most straight-forward approach is to construct a feature vector from the low-level observables of each event and then train a classifier to distinguish pairs of feature vectors corresponding to signal events (i.e.\ simulated with fixed ALP mass and lifetime) from those corresponding to background events (i.e.\ simulated with randomly varying ALP mass and lifetime). The feature vectors for each pair of events are concatenated, and the network is trained to minimize the binary cross-entropy. The classification score for a pair of events $(i,j)$ can then be interpreted as a compatibility measure $C_{ij}$. We call this approach Event Compatibility based on Observables (ECO) hunt, as we base our compatibility assessment on low-level observable.

Alternatively, since we already know that the compatibility of events eventually depends on the compatibility of the corresponding model parameters, we can attempt to reconstruct the posterior probability of the model parameters for each event and provide these as input to a classifier. While this second option is technically more involved, it explicitly makes use of our physical understanding of what defines signal events in order to provide the most relevant information to the algorithm. Another key advantage of extracting the posteriors for each event is that they can be used later for model parameter inference. We will call this approach Events with Posterior Overlap (EPO) hunt, as we are searching for an excess of events whose posteriors cover the same regions of parameter space. 

The posterior extraction can be carried out in different ways, using either classical parameter inference techniques or machine learning. For example, it was shown in Ref.~\cite{Morandini:2023pwj} that accurate posteriors can be obtained in our setup using conditional invertible neural networks. Here we choose a simpler network architecture and employ a classifier to infer the likelihood-to-evidence ratio following Refs.~\cite{Hermans:2019ioj,Miller:2020hua}. We perform (marginalized) neural ratio estimation with classifiers trained on correct and incorrect pairings of data and model parameters. The score of a classifier trained with binary cross-entropy as loss can be converted into the likelihood-to-evidence ratio. By normalizing the probability distribution, we finally get the (marginalized) posterior.

The EPO hunt approach might appear complicated, but it is directly inspired by conventional bump hunting, where one would extract a high-level observable that correlates strongly with the model parameters. To determine whether several events correspond to the same signal hypothesis, however, one needs to know the experimental resolution in terms of the high-level observable, i.e.\ one needs to have an explicit signal template. In our approach, on the other hand, the posterior probability already carries all relevant information, since its maximum corresponds to the best-fit model parameters, while its width reflects the experimental uncertainties.

How to extract model posteriors (or, more generally, likelihood-to-evidence ratios) using ML techniques has been discussed extensively in the literature. It has been shown that classifiers are well suited for this task~\cite{Miller:2022shs,Cranmer:2015bka,Cole:2021gwr,Brehmer:2019xox,Baldi:2016fzo}, since they are comparably easy to train and provide stable results. We note, however, that many alternative approaches exist, using for example normalising flows, conditional invertible neural networks or diffusion models. We will not review the technical details here, but instead refer the reader to \cite{Heimel:2023ngj,Rezende:2015ocs,BayesFlow,papamakarios2021normalizing,Shmakov:2023kjj}. For more literature on this and other ML topics, see \cite{Feickert:2021ajf}.

\begin{figure*}
    \centering
    \includegraphics[width=\textwidth]{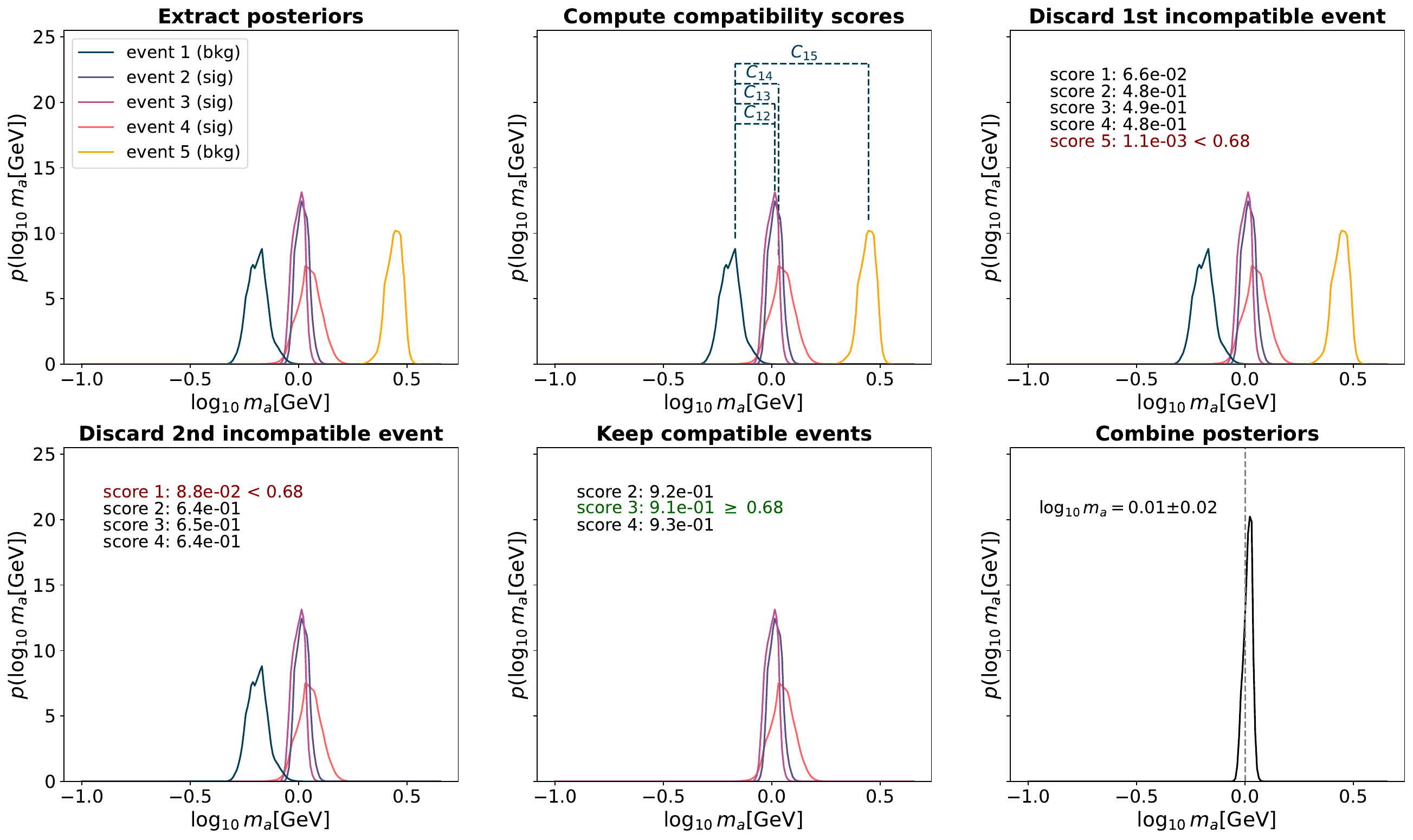}
    \caption{Example analysis procedure with 5 observed events of which 3 are signals and 2 background. The 3 signal events are correctly identified by our algorithm. In the bottom right plot, the posterior of the three signal events combined is shown. The example considered here refers to an ALP mass of $\SI{1}{\giga\electronvolt}$ and the detector setup with poor resolution for angles and energy.}
    \label{fig:procedure_ok}
\end{figure*}

In the case of the ECO hunt, the relevant combinations of low-level observables are constructed internally by optimizing the network weights in order to maximise the accuracy of the network. In the case of the EPO hunt, we directly go to the model parameter space instead of designing observables informative about the model parameters. 

In the end we have three different classifier algorithms:
\begin{enumerate}
    \item a classifier extracting the mass posteriors from the low-level features;
    \item a classifier establishing the compatibility of events based on the low-level observables (ECO hunt);
    \item a classifier establishing the compatibility of events based on the posterior obtained from classifier 1 (EPO hunt).
\end{enumerate}
For details on the network architecture and hyperparameters, as well as the performance validation and coverage tests, we refer to \cref{sec:appendix_architecture}.

\subsection{Analysis strategy}

In the following we will discuss a simplified but realistic data analysis strategy. The discussion focuses on the case of low number of events collected, where the distinction of signal and background is most critical. The data analysis proceeds in three steps:
\begin{enumerate}
    \item Determine whether the background-only hypothesis explains the data.
    \item If the background only hypothesis is rejected, separate the signal and background events.
    \item Infer the signal properties.
\end{enumerate}

Each of these steps will be discussed in a separate section, where we will assess the performance of the different approaches discussed above using appropriate measures. It is worth mentioning that each step could also be carried out separately with a different approach, for example a conventional bump hunt. However, we will see that all the steps can be addressed with our ML approaches with comparable or better performance.

To conclude this section, we visualise the whole analysis procedure for the case of the EPO hunt approach in \cref{fig:procedure_ok}. In this example, the experiment observes five events and their posteriors for the ALP mass are reconstructed using an appropriately trained classifier (first panel). Already by visual inspection, we are able to tell that event 1 is different from the others. This is quantified by computing  compatibility scores, which will be introduced in detail below. This is done first separately for each pair of events (second panel) and then averaged over all pairings (third panel). Here we see that the score of event 1 is indeed lower than our compatibility threshold (set here to 0.68) and hence discard it.

We repeat the procedure with the four remaining events in the fourth panel. After the cleaning procedure, one event is slightly below our compatibility threshold and is therefore removed. We repeat again the cleaning procedure with the three remaining events in the fifth panel. Now all events have a score well above our compatibility threshold. We then combine them to get the combined posterior of the three signal events (and hence an estimate of the ALP mass) in the sixth panel. In this example, the signal events have been correctly identified and the true ALP mass of $m_a = 1 \, \mathrm{GeV}$ was accurately reconstructed. Of course, the procedure will not always work this well, making it necessary to discuss in detail possible performance measures.

\section{Rejecting the background-only hypothesis}
\label{sec:sig_detection}

\subsection{General approach}

Once data has been collected, the first question of interest is whether the data can be described by the background-only hypothesis, or whether there is room for a signal.  This question needs to be answered before any potential signal can be studied in more detail. For counting experiments, an answer can be provided by comparing the number of observed events with the background expectation. More generally, one performs a goodness-of-fit hypothesis test by constructing a test statistic (TS) with known distribution under the background hypothesis and determines the probability of the measured value.

While it is not necessary to have an explicit signal model in order to reject the background-only hypothesis, the construction of the TS may be inspired by the expected difference between signal and background. In other words, to be sensitive to a possible signal, the distribution of the TS should change significantly in the presence of a signal. Here we are interested in signals that are more localised than the background, in the sense that signal events are more similar to each other than background events. Our strategy will therefore be to look for a ``clustering'' of events that is unexpected under the background-only hypothesis. In the simplest case, this is nothing other than the classic bump hunt, where the similarity of signal events is revealed by studying the data in terms of a high-level observable and searching for a localised excess. Similarly, for our classifiers, a useful TS turns out to be the number of compatible events. Let us clarify how this selection algorithm works.

For each pair of events $i$ and $j\neq i$ the classifiers return a compatibility score $C_{ij}\in[0,1]$, where a score of 1 means that the events are identified as coming from the same process. On the other hand, a score of 0 means that the two events are incompatible, even when considering the uncertainties. Signal events are expected to have high compatibility with each other, while background events should have lower compatibility with each other and with signal events. This is analogous to signal events falling into the same bin in a conventional bump hunt, while background events are more broadly distributed. Of course, in both cases nothing prevents a background event to happen close to a signal event, making a perfect separation impossible.

In the case of $n_\text{obs}$ events, we thus obtain $n_\text{obs}(n_\text{obs}-1)/2$ compatibility scores $C_{ij}$.  Once the pairwise compatibility score is established, we clean our sample to contain only the compatible events as follows:
\begin{enumerate}
    \item Evaluate the pairwise compatibility scores $C_{ij}$ with ECO hunt or EPO hunt.
    \item Average them as $s_i=\frac{1}{n_\text{obs}-1}\sum_{j\neq i}C_{ij}$.
    \item Identify the events with the lowest $s_i$.
    \item Compare this value of $s_i$ with a chosen threshold: if the value is above the threshold, conclude. If it is below, discard the event and repeat the previous steps with $n_{\text{obs}}$ reduced by one. 
    \item The number of events retained at the end of the process is used as TS.
\end{enumerate}
By discarding the event with the lowest compatibility, we remove the most background-like event. As a consequence of this discarding, the compatibility between the remaining events increases. A permissive threshold close to $0$ will retain most events and is therefore analogous to using only a few large bins in the case of the conventional bump hunt. A restrictive threshold close to 1 will discard events unless they are extremely similar to each other, analogous to choosing many narrow bins in the usual bump hunt. We find that the optimal choice of the threshold depends only slightly on the number of observed events, the signal mass or the detector resolution, see appendix~\ref{sec:appendix_bkgrej} for further details. 

In principle the threshold could be optimised separately for each signal hypothesis and detector setup. However, in order to perform a first exploratory study in a model independent way, it is convenient to keep the threshold fixed to a value that performs reasonably well for all cases. In the following, we will therefore consider a threshold of 0.68. We have chosen this value as the one that allows the best performance on all combinations of benchmark masses and detector resolutions.

\begin{figure*}
    \centering
    \includegraphics[width=\textwidth]{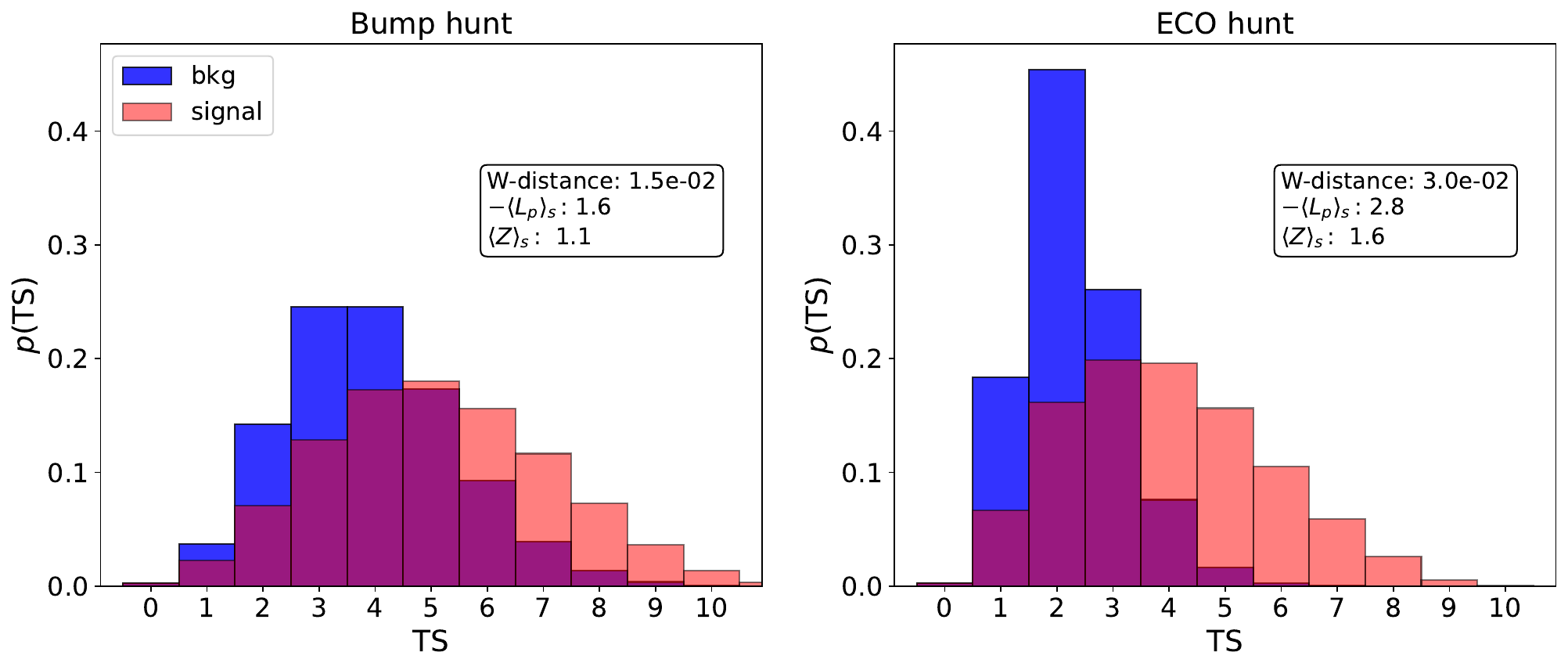}
    \caption{Distribution of the TS for $\mu_b = 6$ (blue) and for $\mu_b=2$, $\mu_s=4$ (red) for the conventional bump hunting (\textit{left}) and ECO hunt (\textit{right}). We consider the case of good detector resolution and assume an ALP mass of $m_a = 1 \, \mathrm{GeV}$ for the signal simulation. The distribution for the case of EPO hunt is very similar to the one of ECO hunt.} 
    \label{fig:countingTS}
\end{figure*}

To summarize, we use the following definition of the TS:
\begin{equation*}
 \text{TS} = \begin{cases} 
 \text{Largest count in single bin} & (\text{bump hunt}) \\
 \text{Number of compatible events} & (\text{ECO \& EPO})\; .   
 \end{cases}
\end{equation*}

As an example we show the distributions of the TS for fixed background and signal in \cref{fig:countingTS} for the case of poor detector resolution. The blue bars correspond to the case that only background is present (with $\mu_b = 6$), while the red bars correspond to a background normalization strength of $\mu_b=2$ and a signal with $\mu_s = 4$ arising from an ALP with $m_a = \SI{1}{\giga\electronvolt}$. In the left panel we show the distribution for a conventional bump hunt with 11 bins, while in the right panel we show the result for the ECO hunt with threshold set to 0.68 \footnote{Note that both the number of bins and the threshold could be optimised further to maximise the sensitivity for this particular setting. We have checked that doing so does not change our conclusions.}.  The EPO hunt gives very similar results to the ECO hunt and is not shown.

The number of events left after the cleaning procedure can now be used as the TS for the background-only hypothesis test. If no events are observed, the value of the TS will be set to 0. 
The TS can therefore be directly compared to the largest number of events observed in a single bin in the bump hunt approach. Of course, for either zero or one observed events, the TS is trivial and all approaches yield identical results. But as soon as at least two events are observed, we can expect the various approaches to perform differently depending on how successfully background events can be removed. 

To quantify the performances we consider specific values of background normalization $\mu_b$ and signal strength $\mu_s$ and assume that the number of observed background (signal) events follows a Poisson distribution with expectation value $\mu_b$ ($\mu_s$). Note that neither $\mu_b$ nor $\mu_s$ directly enter our hypothesis test, i.e.\ we assume that the background normalisation is completely unknown and we do not use the signal model in the hypothesis test. The two parameters are needed only to simulate events and evaluate the distribution of the TS. We then scan over different values of the background normalization and signal strength in order to understand which approach performs best in which region. Since we are interested in the regime of low background and signal, we focus our study to values of background normalization $\mu_b\in[0,6]$ and signal strength $\mu_s\in[0,4]$.

For the case of the bump hunt, the background-only hypothesis predicts roughly one event per bin on average. The actual number of events in each bin follows a Poisson distribution, but due to the bin overlap, the observations in adjacent bins are correlated and an accurate analytical treatment is difficult. Nevertheless, it is intuitively plausible that the majority of simulated realisations will have more than one event in at least one bin, leading to the test statistic peaking around 3. For the case of signal+background, we expect the majority of signal events to end up in the same bin, such that the test statistic takes larger values on average. Nevertheless, the separation between the two cases is not very large, making it difficult to construct a powerful hypothesis test.

For the case of the ECO hunt (right panel), we find that the two distributions become more distinct, because the average value of the TS is shifted to smaller values for the background-only case and to larger values for the background+signal case. Indeed, the value of the TS is found to be very tightly correlated with the number of signal events (with a correlation coefficient of 99.7\%), indicating that the ECO hunt successfully identifies the signal events in most realizations.

\subsection{Performance quantification}\label{sec:performance_sensitivity}

While \cref{fig:countingTS} provides a qualitative proof that the ECO hunt can outperform the conventional bump hunt, we want to compare the respective performances in a quantitative way. In the case of a continuous TS one could for example determine the power of the hypothesis test for fixed significance $\alpha$, for example $\alpha=0.95$. For a discrete TS, however, the power is not a continuous function of the significance and hence the result can depend sensitively on the specific choice of $\alpha$. In the following, we will therefore introduce a number of alternative performance measures.

As discussed above, to obtain a powerful hypothesis test we would like the distribution of the TS to look as different as possible for the case of background-only compared to the case of background+signal. A common measure of how different two distributions are is the Wasserstein distance or Earth-mover distance \cite{10.1023/A:1026543900054}. For the example considered in \cref{fig:countingTS}, this measure takes the value 0.015 for the bump hunt and 0.030 for the ECO hunt, confirming the better performance of the latter. While this separation measure achieves the goal of comparing the different methods without the need for a choice of $\alpha$, the result does not have an intuitive statistical interpretation.

\begin{figure*}
    \centering
    \includegraphics[width=\textwidth]{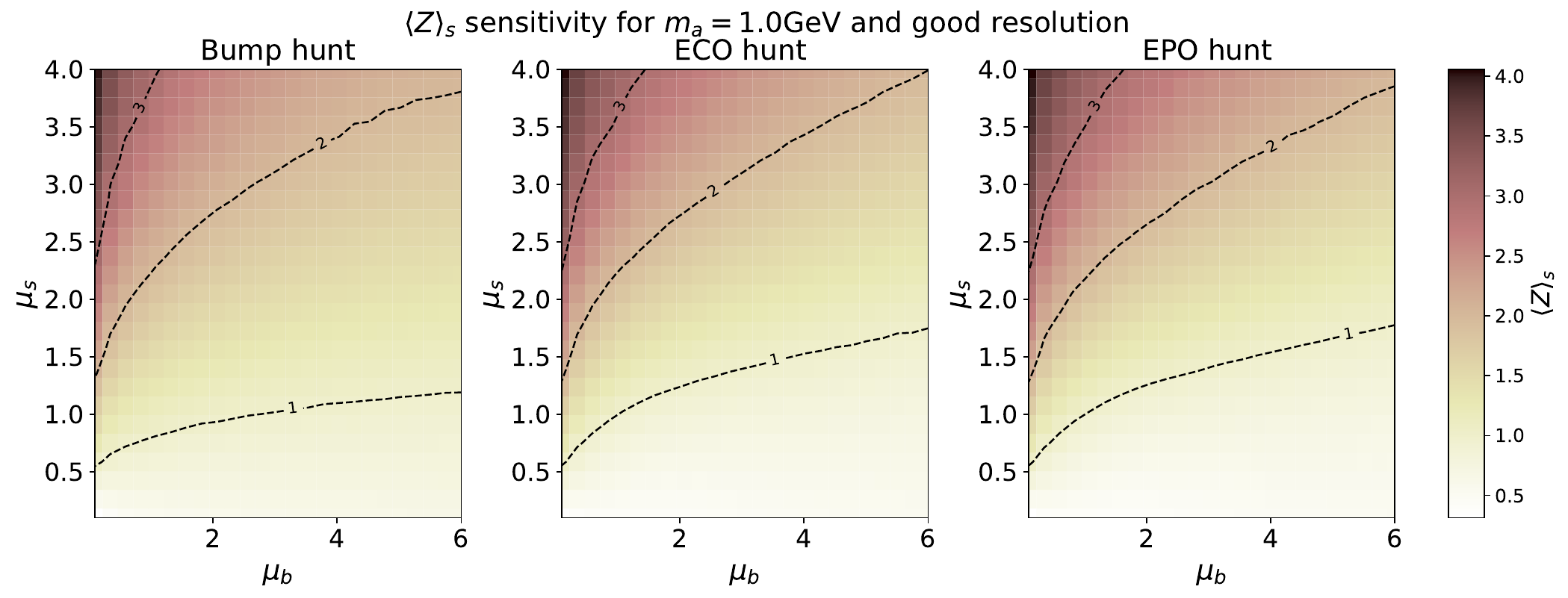}
        \includegraphics[width=\textwidth]{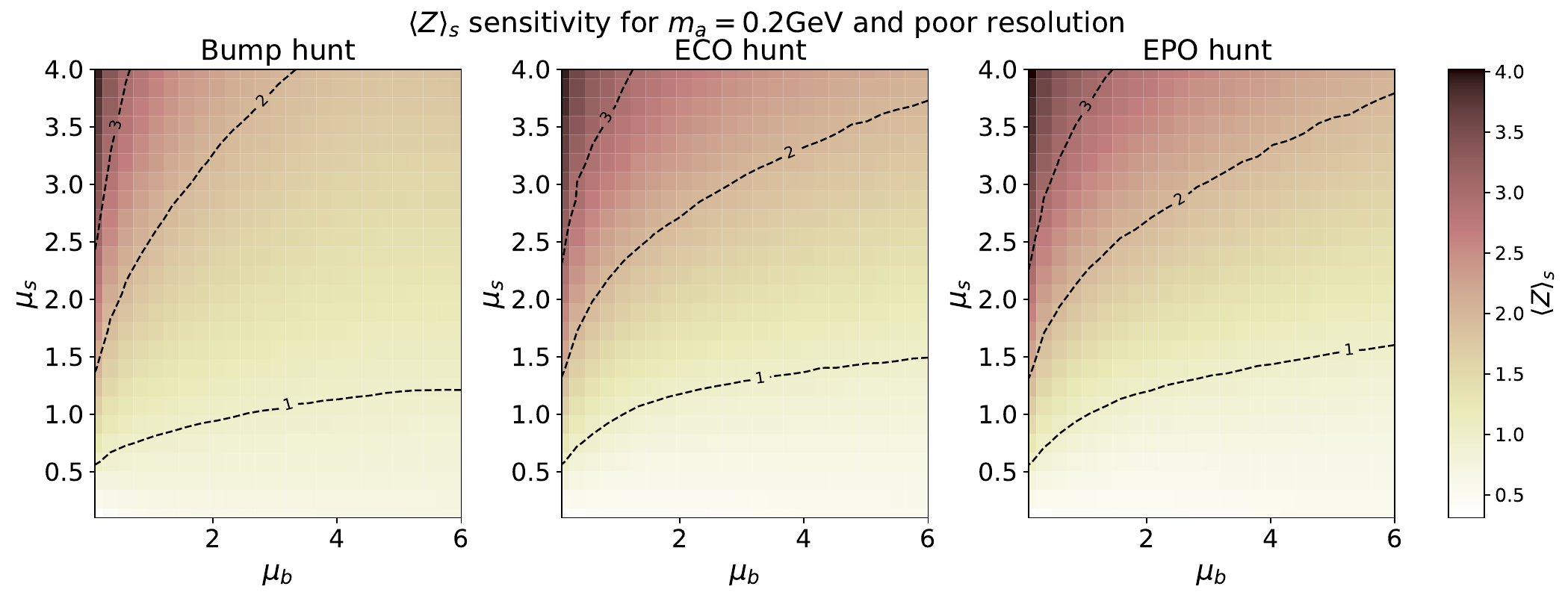}
    \caption{Sensitivity parameter for different cases of signal mass and detector resolution as a function of the background normalisation and signal strength.}\label{fig:sensitivity}
\end{figure*}

To construct a more intuitive performance measure, we define the $p$-value and the $\log p$-value of the background hypothesis as
\begin{align}
    p(l) & = \sum_{m \geq l}^{N_\text{max}} p_b(m) \, , \\
    L_p(l) & = \sum_{m \geq l}^{N_\text{max}} \log p_b(m) \; ,
\end{align}
where $l$ is the observed value of the TS and $p_b(m)$ is the probability to measure $\text{TS} = m$ under the background-only hypothesis. We emphasise that $L_p(l) \neq \log (p(l))$. The $p$-value can also be translated to the equivalent significance $Z(l) = \sqrt{2}\,\mathrm{erf}^{-1}(1-p(l))$, which for a Gaussian distribution corresponds to the number of standard deviations away from the mean.

With these definitions, we can construct two relevant performance measures:
\begin{align}
    -\langle L_p \rangle_s & = - \sum_{l = 0}^{N_\text{max}} p_s(l) L_p(l) \, ,\\
    \langle Z \rangle_s & = \quad \sum_{l = 0}^{N_\text{max}} p_s(l) Z(l) \; ,    
\end{align}
where $p_s$ denotes the probability distribution of the TS for the case of signal + background. In other words, these performance measures correspond to the expectation value of the $\log p$-value and of the significance for the case of signal + background. In both cases, larger values correspond to a larger separation of background and signal + background, i.e.\ a better performance of the method under consideration. For the example considered in \cref{fig:countingTS}, we find $-\langle L_p \rangle_s = 1.6$ ($-\langle L_p \rangle_s = 2.8$) and $\langle Z \rangle_s = 1.1$ ($\langle Z \rangle_s = 1.6$) for the bump hunt (ECO hunt), confirming our qualitative observation that the ECO hunt performs much better than the bump hunt in this particular case. However, as we will show next, the performance measures depend on $\mu_b$ and $\mu_s$, as well as on the detector resolution and the specific signal hypothesis.

\begin{figure*}
    \centering
        \includegraphics[width=\textwidth]{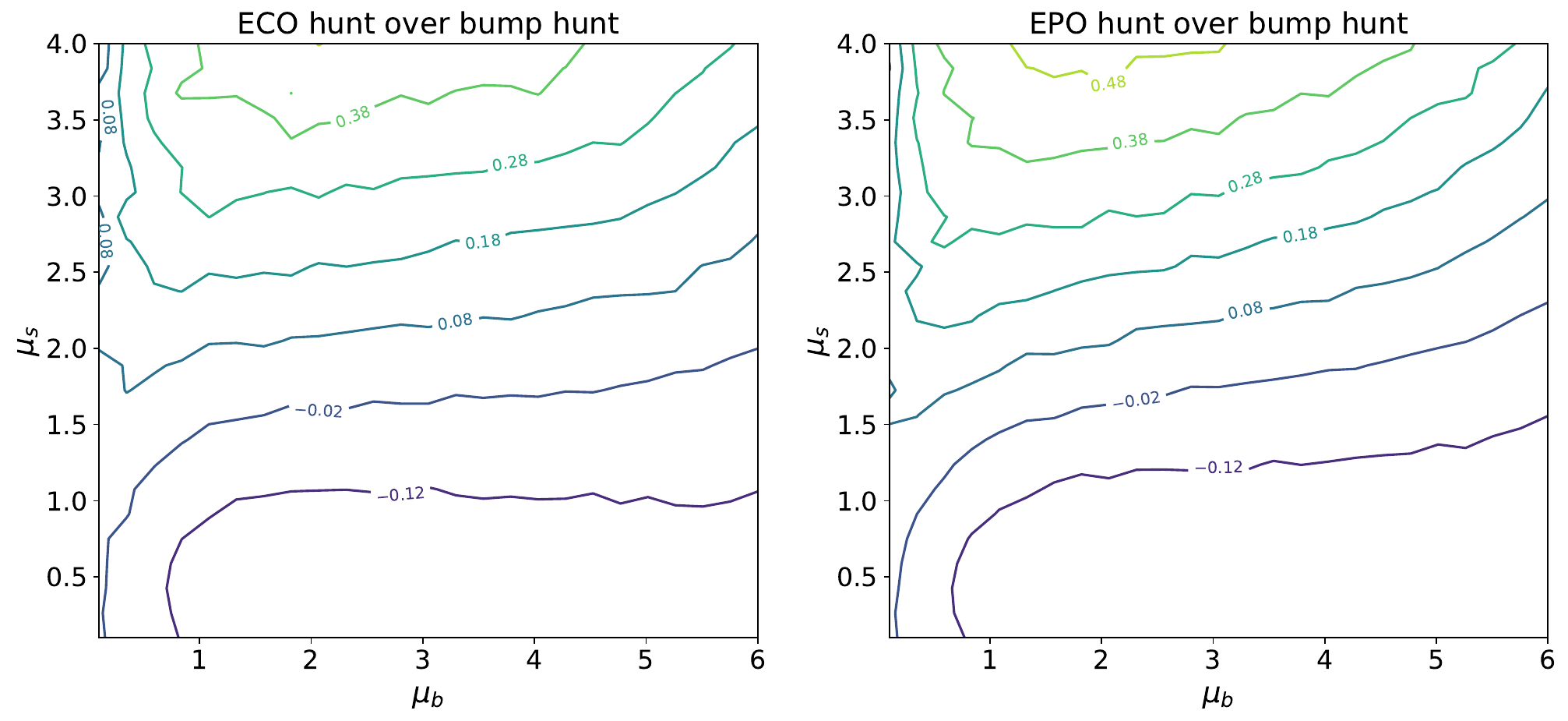}
    \caption{Improvement over the classic bump hunt achieved by the ECO hunt (left panel) and the EPO hunt (right panel) in terms of the sensitivity parameters $\langle Z\rangle_{s}$ as a function of the background normalisation and signal strength. The ALP mass is set to $m_a = 0.2 \, \mathrm{GeV}$ and we consider a detector with poor resolution.}\label{fig:improv}
\end{figure*}

\subsection{Results}

We are now in the position to perform a detailed comparison of the performance of the different methods, i.e.\ conventional bump hunting, hunting in observables (ECO hunt) and hunting in posteriors (EPO hunt), using the discovery sensitivity parameter $\langle Z \rangle_s$. 

In \cref{fig:sensitivity} we visualise the values of the sensitivity parameter for different values of signal strength $\mu_s$ and background normalization $\mu_b$. We focus here on two different cases: a large mass of $\SI{1}{\giga\electronvolt}$ observed by a calorimeter with good resolution and a small mass of $\SI{0.2}{\giga\electronvolt}$ observed by a calorimeter with poor resolution. As shown in Ref.~\cite{Morandini:2023pwj}, these two cases represent respectively a scenario where the standard approach (i.e.\ the bump hunt) should work well and a scenario where the standard approach would have limitations.

In both scenarios, all approaches exhibit roughly the same behaviour, namely that the expected sensitivity (i.e.\ the power to reject the background-only hypothesis) decreases for larger $\mu_b$ and increases for larger $\mu_s$. In the first scenario the performances are very similar regardless of the applied approach. On the contrary, in the case of small mass the performance of the conventional approach decreases visibly, while the performance of the ML-based approaches is almost unchanged. This suggests that the ML approaches are more robust against an increase in experimental uncertainties. This finding will be further corroborated when discussing background events rejection in the following section, where again the larger detector uncertainties do not significantly affect the performance of the ML approaches. 

We can also see in \cref{fig:sensitivity} that the ECO hunt and EPO hunt have very similar performances. This implies that most of the information contained in the event (or at least the information necessary to assert the compatibility of events) is contained in the mass posterior. Conversely, this also suggests that the ECO hunt (which has the full information of the underlying event) is internally performing a model parameter reconstruction in order to assess the event compatibility. The fact that the EPO hunt slightly outperforms the ECO hunt in some cases suggests that the former is easier to train. If optimally trained, the ECO hunt should perform at least as well as the EPO hunt, because it receives the same inputs and has the more expressive architecture.

For the case of small ALP mass and poor resolution, we can visualise the improvement achieved by the two ML approaches compared to the bump hunt by taking the difference between the respective sensitivities. These are visualised in \cref{fig:improv}. As expected, the improvement is very modest if both $\mu_s$ and $\mu_b$ are small, in which case it will generally not be possible to establish the presence of a signal. On the other hand, in the most interesting region with relatively large $\mu_s$ and $\mu_b$ the sensitivity increases by up to 0.75. This is a substantial improvement, keeping in mind that the naive scaling $Z \propto \mu_s / \sqrt{\mu_b}$ would suggest that we need to double the size of the data set to increase the sensitivity from $2$ to $2.8$. The improvements of the ECO and EPO hunt over the bump hunt follow very similar trends in $\mu_b$ and $\mu_s$, with the EPO hunt giving slightly better values. The fact that the improvement is largest for larger numbers of observed events is expected, as the advantage of the ML approaches resides in the improved ability of removing background events from the sample. This ability will be investigated in more detail in the following section.

\section{Removing background events}
\label{sec:bkg_remove}

\begin{figure*}
    \centering
    \includegraphics[width=0.49\textwidth]{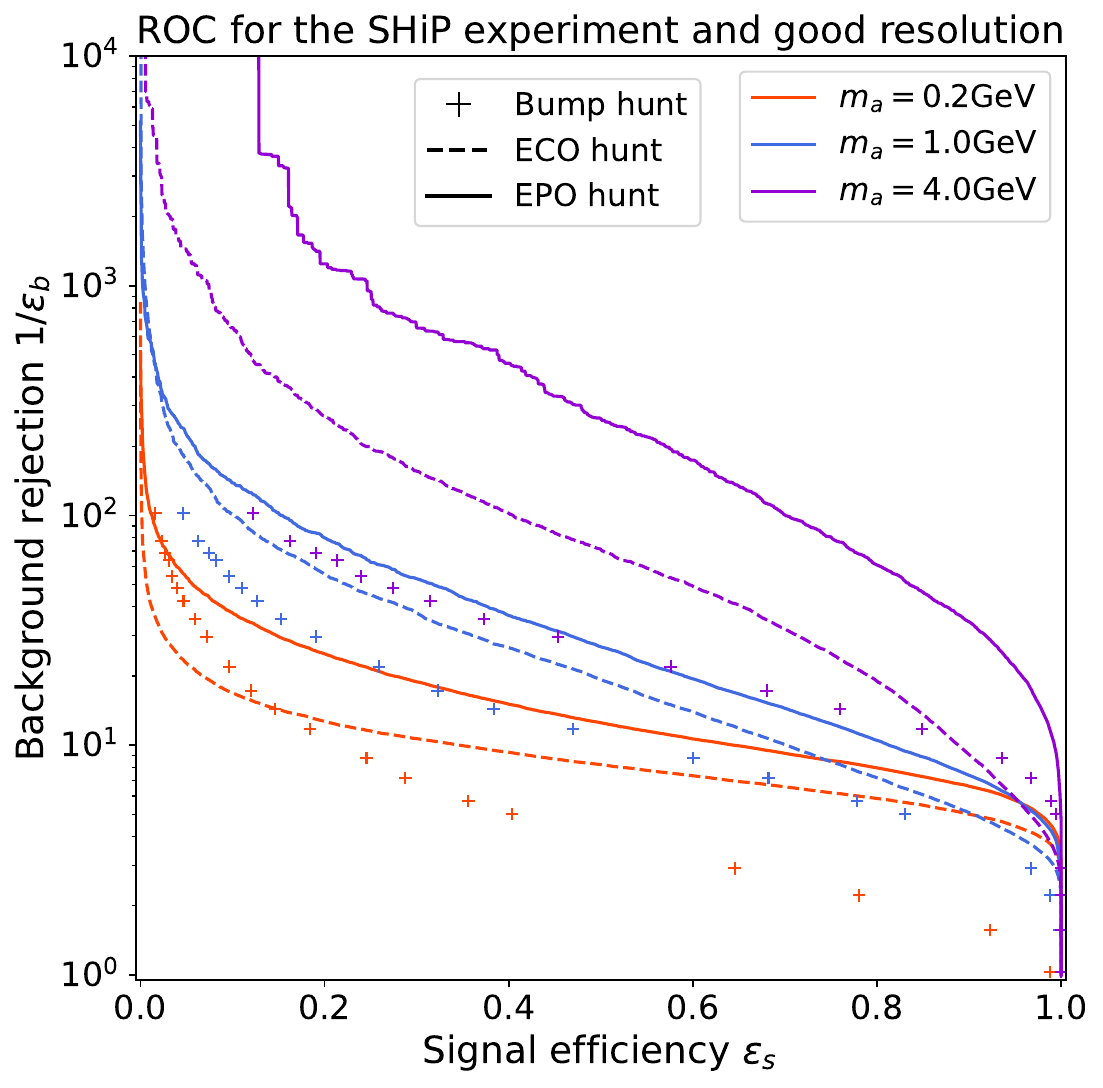} \includegraphics[width=0.49\textwidth]{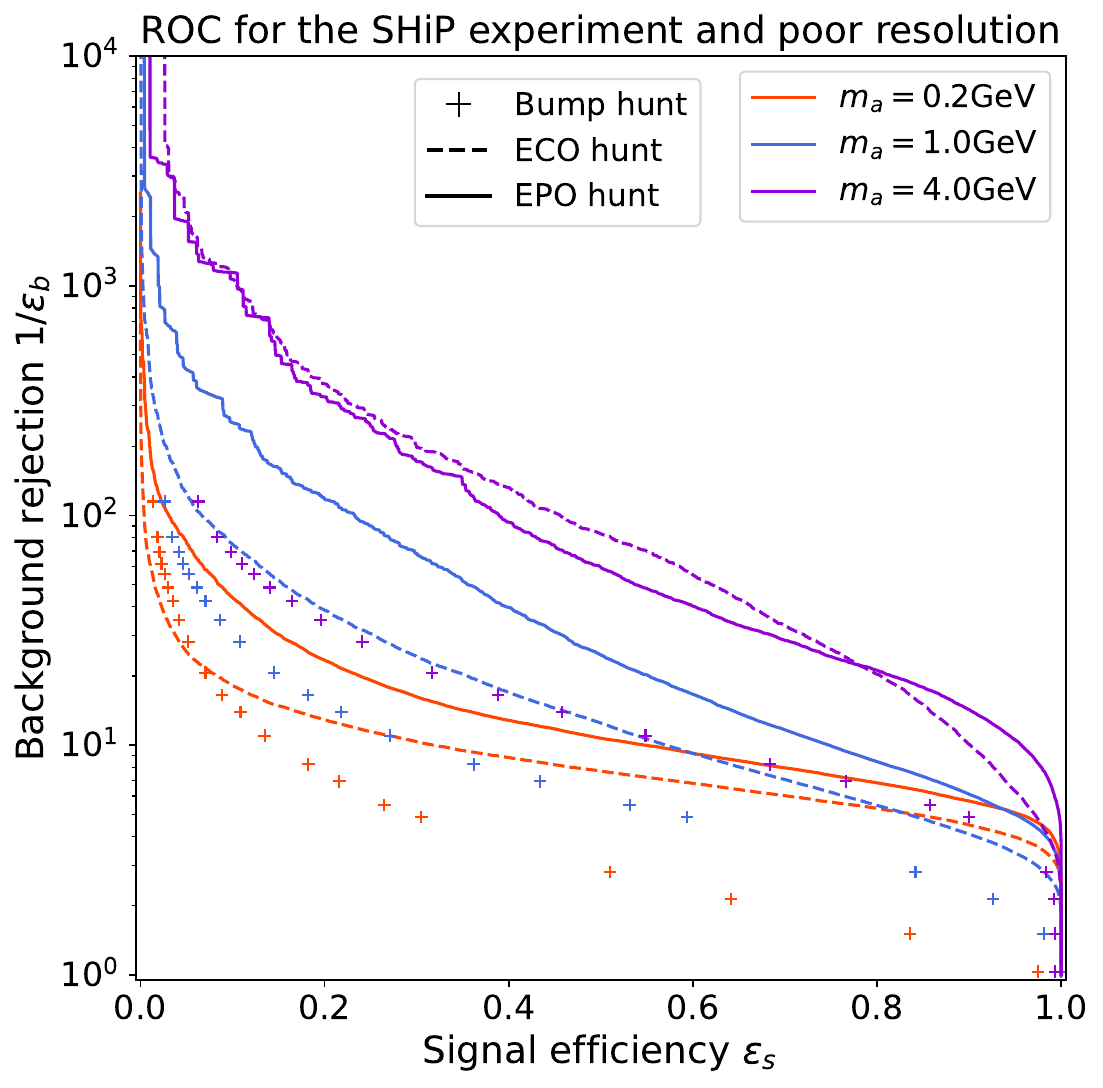}
    \caption{ROC curve and AUC in the case of small (left panel) and large (right panel) uncertainties. The EPO hunt is shown in solid, the ECO hunt in dashed, and the conventional bump hunt is indicated with $+$ markers. As we move from the left to the right of the plot, the bin width and compatibility threshold increase, leading to larger signal efficiency and smaller background rejection.}
    \label{fig:roc}
\end{figure*}

\subsection{Approach and performance quantification}

The hypothesis tests in the previous section relied implicitly on the distinction between signal and background events. This distinction is necessary for the TS distribution to be different under the background-only and signal+background hypothesis. In the present section, we make explicit use of these differences to remove background events from a sample of observed events.

In the ideal case, one would be able to find a signal region (or a signal window in the case of a single high-level observable) such that all signal events fall inside that region and all background events end up outside of it. In practice, there will be a non-zero probability for  background events to fall into the signal region, in particular if the region is large. For a small signal region, on the other hand, signal events can end up outside of it when taking into account the experimental resolution. The optimal size of the signal region therefore balances these two opposite effects, i.e.\ it needs to find a compromise between falsely accepted background events and falsely rejected signal events.

With our ML approaches we face exactly the same issue. If we require a high compatibility threshold in order for events to count as signal, we risk removing signal events. Lowering the threshold for compatibility, on the other hand, increases the chance of keeping background events.

A straightforward way to quantify the performance of the task to correctly identify signal events is to determine the Receiver Operator Characteristic (ROC) curve, i.e.\ the background rejection ($1/\varepsilon_b$) as a function of signal efficiency ($\varepsilon_s$), as well as the corresponding Area Under Curve (AUC). The ROC curve is obtained by varying the threshold value between 0 (all events accepted, $\varepsilon_s=\varepsilon_b=1$) and 1 (all events rejected, $\varepsilon_s=\varepsilon_b=0$).\footnote{Note that in our approach the number of bins in the conventional bump hunt can only take integer values. As a result, we do not actually obtain a continuous ROC curve in this case, but a discrete set of signal acceptances and corresponding background rejections.} These quantities are useful measures of our overall ability to correctly assess the compatibility of two events and identify them as signal.

In the physical scenario we are interested in, however, not all parts of the ROC curve are equally important to us. Since we are interested in scenarios with a small number of signal events, we want to keep as many of them as possible in order to constrain the underlying signal model, i.e.\ we want to focus on the background rejection for large signal acceptance. A cumulative measure like the AUC is therefore too simplistic. 
A more suitable measure of performance is for example given by the background rejection for $\varepsilon_s = 0.8$ or $\varepsilon_s = 0.9$. Clearly, higher background rejection corresponds to better performance. 

\subsection{Results}\label{sec:rej_results}

\sisetup{round-mode=uncertainty,round-precision=1,scientific-notation=true,table-omit-exponent, separate-uncertainty=true, multi-part-units=single}
\begin{table*}[]
    \centering
        \caption{ECO hunt background rejection for fixed signal efficiency and AUC.}
    \label{tab:eff_hunt_obs}
    \begin{tabular}{c C C C C}
    \hline \hline
       resolution  & m_a [\SI{}{\giga\electronvolt}]&\frac{1}{\varepsilon_b}(\varepsilon_s=0.8)&\frac{1}{\varepsilon_b}(\varepsilon_s=0.9)& \text{AUC}\\
       \hline \hline good & 0.2 & \num{5.857886  \pm 0.149635  } & \num{4.999144  \pm 0.084307} & \num{ 0.873346 \pm 0.004703} \\
good & 1.0 & \num{7.208621  \pm 0.633783  } & \num{5.076961  \pm 0.352892} & \num{ 0.918912 \pm 0.005223} \\
good & 4.0 & \num{18.925751 \pm 5.141598  } & \num{9.073154  \pm 1.713430} & \num{ 0.961661 \pm 0.006207} \\
\hline
poor & 0.2 & \num{5.309271  \pm 0.346477  } & \num{4.488659  \pm 0.260081} & \num{ 0.863600 \pm 0.011699} \\
poor & 1.0 & \num{5.469323  \pm 0.361744  } & \num{4.099832  \pm 0.268064} & \num{ 0.891553 \pm 0.007517} \\
poor & 4.0 & \num{20.347858 \pm 10.576270 } & \num{10.066523 \pm 3.911786} & \num{ 0.961212 \pm 0.011950} \\
\hline \hline
    \end{tabular}
\end{table*}

\sisetup{round-mode=uncertainty,round-precision=1,scientific-notation=true,table-omit-exponent, separate-uncertainty=true, multi-part-units=single}
\begin{table*}[]
    \centering
        \caption{EPO hunt background rejection for fixed signal efficiency and AUC.}
    \label{tab:eff_hunt_post}
    \begin{tabular}{c C C C C}
    \hline \hline
       resolution  & m_a [\SI{}{\giga\electronvolt}]&\frac{1}{\varepsilon_b}(\varepsilon_s=0.8)&\frac{1}{\varepsilon_b}(\varepsilon_s=0.9)& \text{AUC}\\
       \hline \hline good & 0.2 & \num{7.963110  \pm 0.252073 } & \num{ 6.627059  \pm 0.193064 } & \num{0.913582  \pm 0.006362} \\
good & 1.0 & \num{10.485804 \pm 0.788886 } & \num{ 7.382669  \pm 0.357716 } & \num{0.941987  \pm 0.006406} \\
good & 4.0 & \num{60.809548 \pm 22.728717} & \num{ 34.375500 \pm 12.088700} & \num{0.987668  \pm 0.004687} \\
\hline
poor & 0.2 & \num{6.848192  \pm 0.278642 } & \num{ 5.710981  \pm 0.193758 } & \num{0.902331  \pm 0.004617} \\
poor & 1.0 & \num{8.471560  \pm 0.734978 } & \num{ 6.107328  \pm 0.253249 } & \num{0.931989  \pm 0.009537} \\
poor & 4.0 & \num{21.163884 \pm 7.649330 } & \num{ 14.268936 \pm 4.181071 } & \num{0.967757  \pm 0.011057} \\
\hline \hline
    \end{tabular}
\end{table*}

We start by considering the ROC curves for the different signal benchmark masses. They are visualised in \cref{fig:roc}, with the left (right) panel corresponding to small (large) uncertainties. In addition to the result of the ECO hunt approach (dashed lines) and EPO hunt approach (solid lines), we also visualise the performances for different bin widths (crosses). 
Different colours correspond to different ALP masses. We find that for all approaches the performance is better for larger ALP masses, consistent with our previous results and the intuition that larger ALP masses lead to larger photon opening angles, which are less affected by the angular resolution.

When comparing the overall performance, we find that the EPO hunt and ECO hunt generally perform similarly well. In most cases we furthermore find a slightly better performance for the EPO hunt (see \cref{tab:eff_hunt_obs,tab:eff_hunt_post} for further details). This finding is somewhat surprising, given that the ECO hunt has access to the full set of low-level observables for each event, while the EPO hunt only knows about the inferred mass posterior. We conclude that this mass posterior compresses the most relevant information into a form that allows for very efficient training.  

When going from the case of good to poor resolution, the performances deteriorate only slightly for the ML-based approaches, while they get substantially worse for the conventional bump hunt. For good resolution, the conventional approach shows similar performances as the ML approaches, and even outperforms the latter in cases, for example for $m_a=\SI{0.2}{\giga\electronvolt}$ and low signal efficiency. For the case of poor resolution, the situation is different, and the ML-based approaches clearly outperform the conventional one.

In \cref{tab:eff_hunt_obs,tab:eff_hunt_post} we also quantify the network uncertainties for the various performance measures, obtained by repeating the training several times with different initializations. We find that these different networks perform differently well for different signal hypotheses, i.e.\ one performs better for small ALP masses, while another performs better for ALP masses. While the resulting uncertainties are non-negligible, they do not qualitatively change our conclusions.

With our fixed threshold of 0.68, we are always in the desirable region of high signal efficiency. For ECO hunt we end up with $\epsilon_s=0.72\pm 0.06$ for the worst case scenario of small mass and poor resolution, while for the best scenario of large mass we have $\epsilon_s=0.94\pm 0.02$. The same threshold for EPO hunt puts us in an even higher signal efficiency region, explicitly we have $\epsilon_s = 0.83\pm0.02$ for the worst case and $\epsilon_s = 0.99\pm0.01$ for the best case scenario.

While in the previous section we saw only a moderate improvement of the ECO and EPO hunt over the standard bump hunt, the situation changes when we consider the rejection of background events. In particular, EPO hunt always outperforms the conventional approach, even for the case of good resolution. Overall, both of our ML approaches show good performance in the high signal efficiency region for our fixed threshold choice. This will allow us to perform parameter inference with the selected events in the next section.

\section{Parameter inference}
\label{sec:inference}

\subsection{Approach and performance quantification}

We have now established how we can use our approach to determine the presence of signal and to remove background events from a sample of observed events. The remaining events, identified as coming from a common signal, can be used to infer the parameters of the model. In Ref.~\cite{Morandini:2023pwj} we already studied one possible way to achieve this goal using conditional invertible neural networks trained on sets of events. Here we consider an alternative possibility, namely to use the parameter posteriors for individual events, which we constructed in the context of the EPO hunt approach. By combining the single-event posteriors, we can extract the posterior from multiple observed signal events.

To perform the combination, we note that the likelihood from a group of $n$ independent events is just the product of the single-event likelihoods:
\begin{equation}
    \mathcal{L}(\mathbf{x}_1,,\mathbf{x}_n|\theta)=\prod_{i=1}^{n} \mathcal{L}(\mathbf{x}_i|\theta),
\end{equation}
where $\mathbf{x}_i$ is the feature vector of the $i$th event and $\theta$ is the model parameter we are interested in. In general there can be multiple model parameters, but in our case we are focusing on the marginalised posterior on the mass, so we will restrict to a single model parameter for this discussion.
The Bayesian posterior for several events is then given by
\begin{equation}\label{eq:post_n}
    p(\theta|\mathbf{x}_1,,\mathbf{x}_n)=\frac{\pi(\theta) \prod_{i=1}^n\mathcal{L}(\mathbf{x}_i|\theta)}{p}
\end{equation}
with the prior $\pi(\theta)$ and the Bayesian evidence $p$, which provides the normalisation for the posterior.

We emphasise that the Bayesian posterior for several events is not simply the product of the posteriors for individual events:
\begin{equation}\label{eq:post_n}
    p(\theta|\mathbf{x}_1,,\mathbf{x}_n) \neq \prod_{i=1}^n \frac{\mathcal{L}(\mathbf{x}_i|\theta)\pi(\theta)}{p_i}
\end{equation}
both because $\pi(\theta) \neq \pi^n(\theta)$ and because $p \neq \prod p_i$. However, given the posterior of an individual event, we can directly extract the likelihood-to-evidence ratio $r(\mathbf{x}_i|\theta)=\frac{\mathcal{L}(\mathbf{x}_i|\theta)}{p(\mathbf{x}_i)}$ by dividing the posterior by the prior. In fact, $r(\mathbf{x}_i, \theta)$ is actually the quantity provided by the classifier used in the EPO hunt. These ratios can then be multiplied together and with the prior $\pi(\theta)$ to obtain a quantity that differs from the actual posterior only by a multiplicative factor:
\begin{equation}\label{eq:post_n}
    p(\theta|\mathbf{x}_1,,\mathbf{x}_n) \propto \prod_{i=1}^n r(\mathbf{x}_i|\theta) \pi(\theta) \; .
\end{equation}
The missing factor can be obtained by numerically integrating the right-hand side and making use of the fact that the left-hand side is normalised to unity. In our case, where we want to obtain the marginalised posterior for the mass, the normalization requires only a 1-dimensional integration at very little computational cost. No further algorithm training or assumptions are necessary to extract the model parameter.

While this procedure is very simple, it should be kept in mind that every evaluated posterior carries some error. The larger the number of events we want to combine, the larger the combined error could become. For this reason it may be advantageous (in the case of a large event sample) to train an inference algorithm directly on the whole sample, rather than combining the posteriors of the single events. However, for the case of classifiers and likelihoods, it has been shown that combining multiple events before training the classifier does not lead to a better performance \cite{Nachman:2021yvi}. In order to check the reliability of the posterior combination, we will check the coverage plots as discussed in \cref{sec:appendix_architecture}.

Another possible concern is that the procedure outlined above gives undesired results in the case that a signal event is incorrectly discarded or a background event is accepted. In the following we will consider these cases and show that the procedure is rather robust and leads to reliable results.

\subsection{Results}

\begin{figure}
    \centering
    \includegraphics[width=0.5\textwidth]{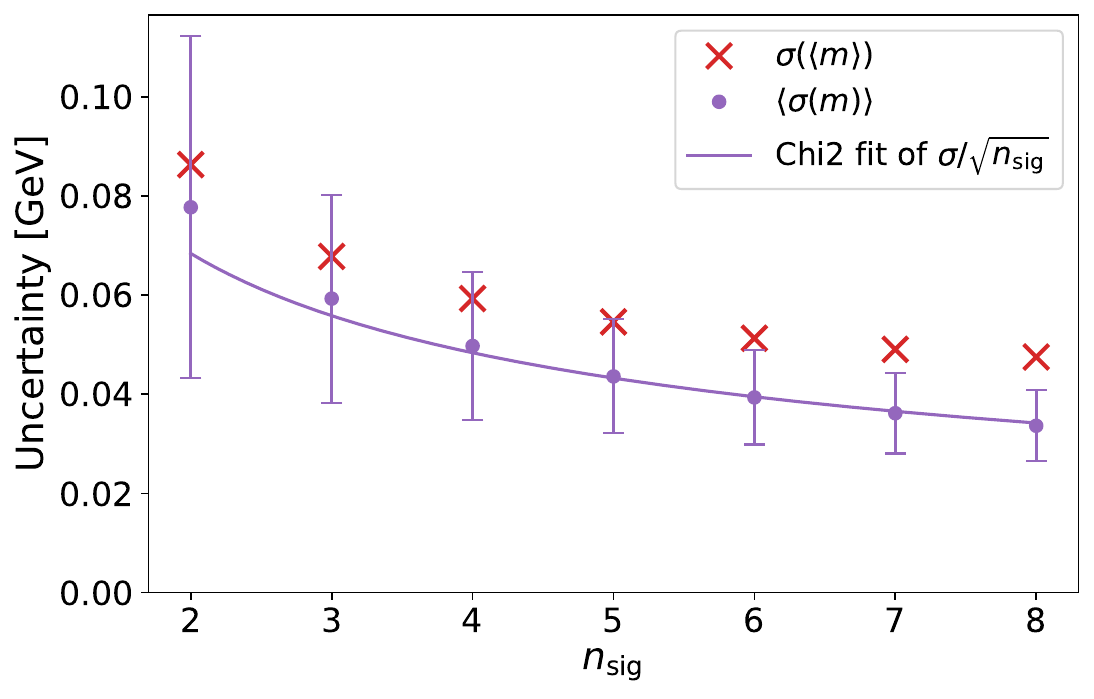}
    \caption{Behaviour of the uncertainty on the inferred ALP mass for increasing number of signal events. We show both the spread of the average prediction and the average posterior width. The predicted scaling would follow $1/\sqrt{n_\text{sig}}$, indicated by the line.}
    \label{fig:Nscaling}
\end{figure}

We have already shown an example of combining signal events in \cref{fig:procedure_ok}. This particular example was chosen such that all the signal events are correctly identified. If this is the case, the posterior behaves as expected, i.e.\  both the width of the posterior and the standard deviation of the posterior peak $\sigma(\log_{10}\hat m)$ for multiple realisations of the experiment scale as $(n_\text{sig})^{-1/2}$ when increasing the number of observed events $n_{sig}$, see \cref{fig:Nscaling}. We conclude that the combined posterior has approximately the correct coverage, at least for $n_\text{sig} \leq 8$.

\begin{figure}
    \centering
    \includegraphics[width=0.5\textwidth]{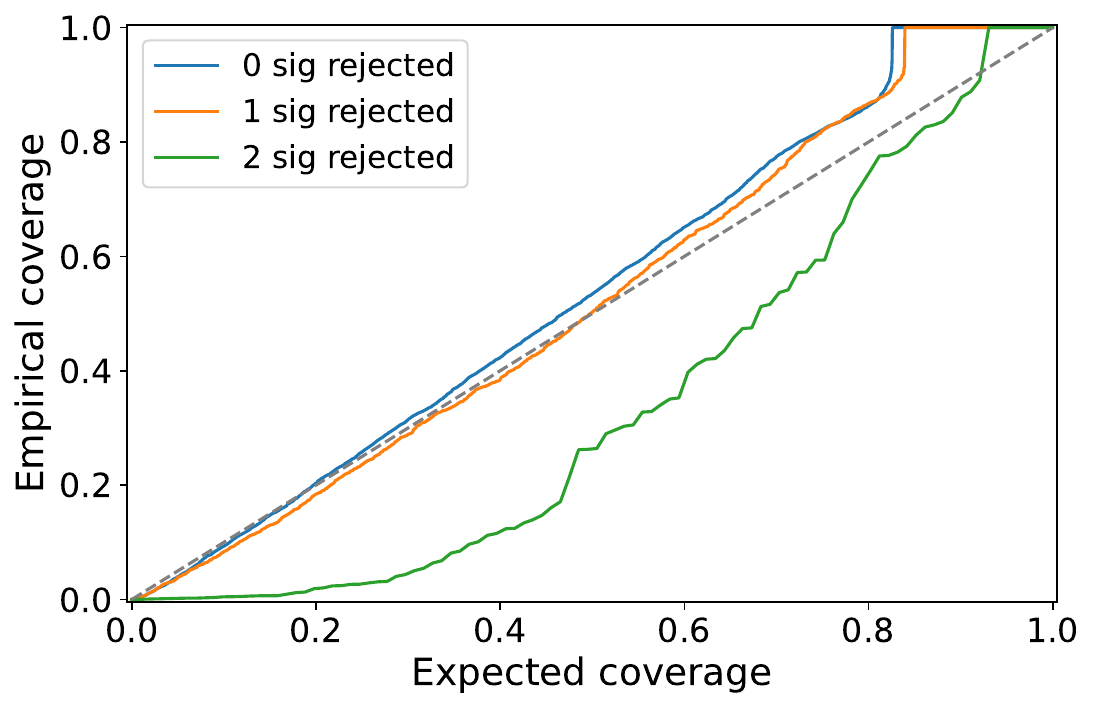}
    \includegraphics[width=0.5\textwidth]{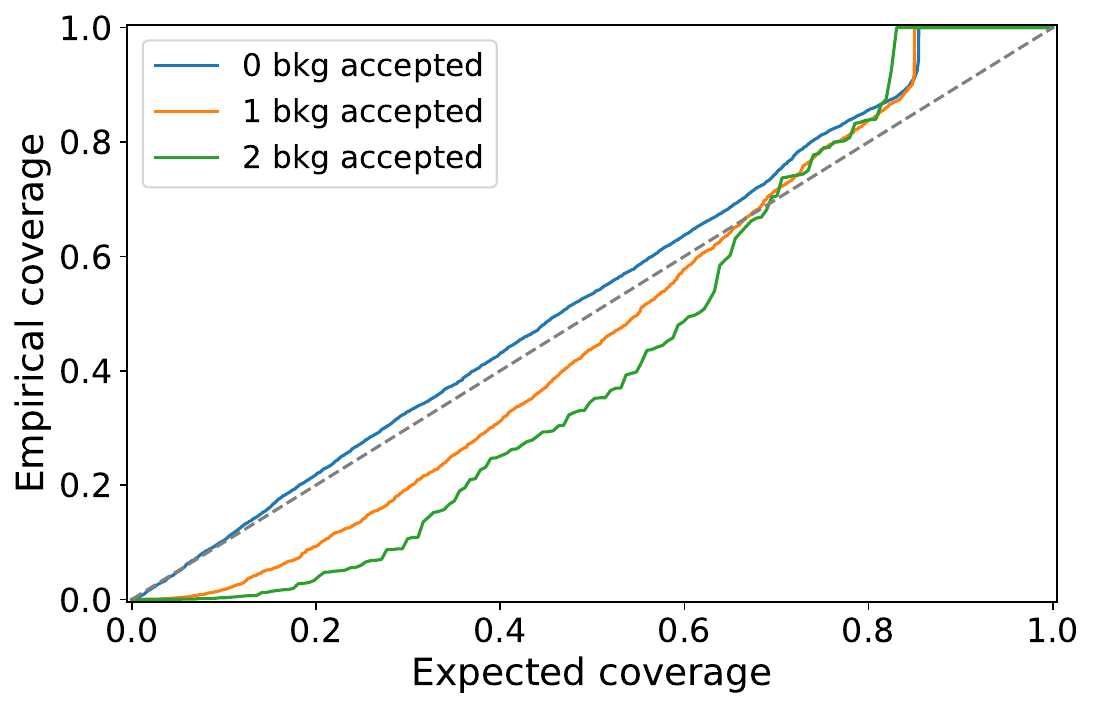}
    \caption{Coverage test when discarding signal events (top) and when accepting background events (bottom). The scenarios considered reflect the scenarios in \cref{fig:procedure_notok}, so four signal events plus one background event (top) and three signal events plus two background events (bottom).}
    \label{fig:coverage_sig}
\end{figure}

This is confirmed in \cref{fig:coverage_sig}, where the blue lines in the two panels show the coverage for the case of three and four correctly identified signal events. In the coverage plots we can understand whether the inferred posteriors are reliable or not. Empirical coverages larger than expected indicate that the uncertainties on the model parameters are larger (i.e.\ the corresponding posteriors wider) than necessary. On the contrary, empirical coverages smaller than expected can be due to either a bias in the parameter prediction or underestimated uncertainties. This is exactly what happens when signal events are incorrectly rejected or background events are incorrectly accepted.

\begin{figure*}
    \centering
    \includegraphics[width=\textwidth]{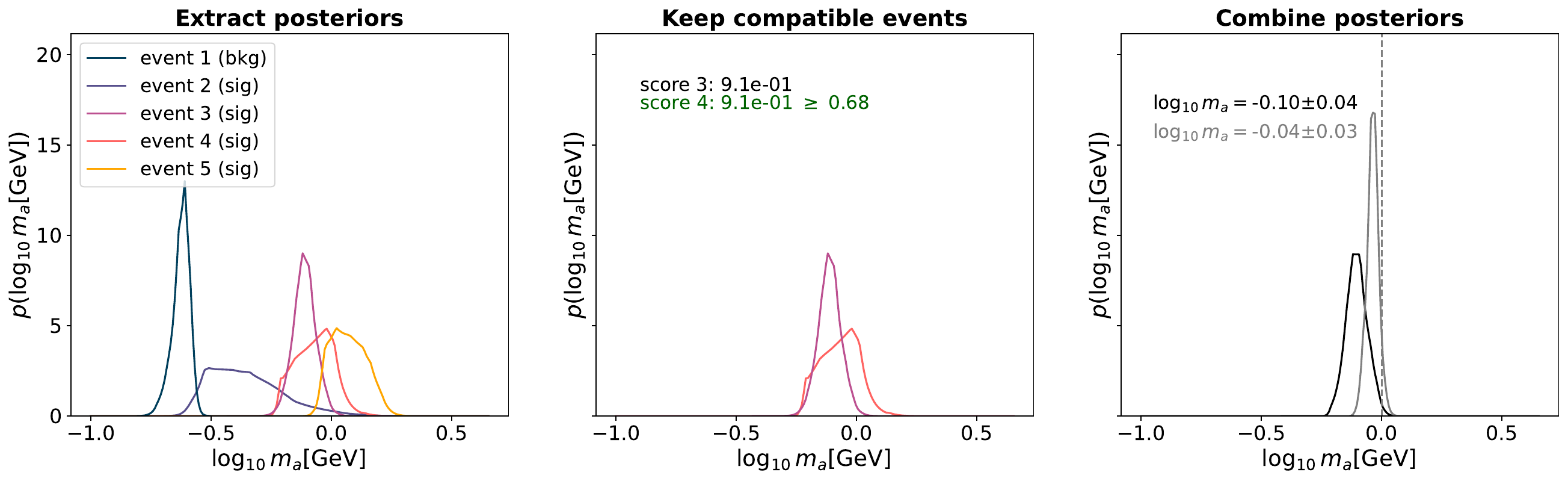}
    \includegraphics[width=\textwidth]{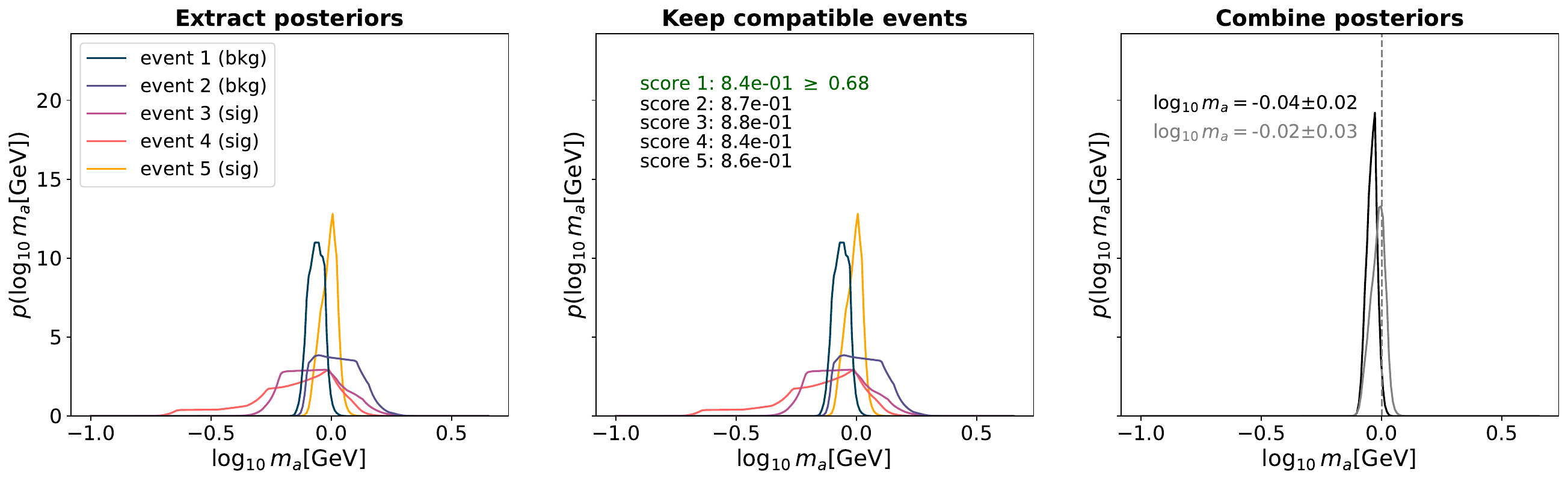}
    \caption{Example analysis procedure with 5 observed events when two signal events are rejected (\textit{top}) or two background events are accepted (\textit{bottom}). In the plots on the right the combination of selected events in shown in black and the posterior of the true signal events is shown in grey.}
    \label{fig:procedure_notok}
    \vspace{-3mm}
\end{figure*}

However, from the results presented in \cref{sec:rej_results}, we know that this does not always happen. We therefore need to consider also what happens when background events are not rejected and when signal events are not accepted. Examples for these two cases are shown in \cref{fig:procedure_notok}.  In this section we always consider an ALP mass of $\SI{1}{\giga\electronvolt}$ and the detector setup with poor resolution, but our conclusions extend qualitatively to all the benchmarks.

In the top row of \cref{fig:procedure_notok} we show what happens if two of the four signal events are incorrectly discarded. Discarding signal events has two effects: First, the resulting posterior will be broader than it would be if all signal events were included. Second, because outlier events tend to be removed first, the peak of the posterior will be biased and therefore subject to larger fluctuations. Both effects can be seen in the top-right panel of \cref{fig:procedure_notok}, where we compare the posterior obtained combining the selected events (in black) to the one obtained by combining all true signal events (in grey). In general, the second effect dominates, i.e.\ for repeated experiments the peak of the posterior exhibits larger fluctuations than expected from the width of the posterior. This corresponds to an undercoverage, because the true value of the ALP mass lies inside the 68\% credible region in less than 68\% of the realisations. This expectation is confirmed in \cref{fig:coverage_sig}, where we show the coverage in case of incorrectly rejected signal events.

The case where background events are incorrectly accepted as signal is visualised in the second row of \cref{fig:procedure_notok}. In this specific case, there is no threshold that could reject the background events while retaining all signal events.
Including additional events leads to a narrower and shifted posterior. Clearly, this case also corresponds to undercoverage, i.e.\ the small width of the posterior does not accurately reflect the true uncertainty of the inferred ALP mass. This is shown in the bottom panel of \cref{fig:coverage_sig}. 

Nevertheless, we observe from \cref{fig:coverage_sig} that the issue of undercoverage is most severe only for relatively low coverage. If we are interested in inferring the 68\% credible interval for the ALP mass, our results are not significantly affected by either rejecting signal events or accepting background events. The combination of single-event posteriors obtained within the EPO hunt approach is therefore expected to provide an accurate characterisation of the underlying signal.

\section{Conclusions}
\label{sec:conclusions}

Simulation-based inference methods offer the potential to directly analyse experimental data using a large number of low-level features instead of a few high-level observables. In the current work we have shown that this approach can be used to search for an excess over expected background using only low-level features. The fundamental idea is to identify events that are more similar to each other than to the background events. This similarity can be understood as a commonality in the model parameters inferred from each event. In other words, one hunts for an excess in model parameter space.

As an explicit example, we have considered the search for axion-like particles decaying into two photons at the future proton beam-dump experiment SHiP. For this search traditional bump hunting reaches its limits, because it is difficult to accurately measure the final state and reconstruct the decay vertex. As a result, it is suboptimal to use the diphoton mass as high-level observable to search for an excess.

To address this challenge, we have developed two different approaches based on classifiers: ECO hunt and EPO hunt. While the former directly uses low-level observables to search for similarity between events, the latter first reconstructs the posterior of the ALP mass for further analysis. We have shown that both approaches perform similarly well when testing the background-only hypothesis and when separating signal and background, suggesting that the ALP mass posterior carries all the relevant information. Compared to the ECO hunt, the EPO hunt has the advantage that the individual posteriors of the events that remain after background-signal separation can directly be used for model parameter inference by combining them into a single posterior of a common signal. 

Fig. \ref{fig:sensitivity} shows that both approaches are promising alternatives to traditional bump hunting and increase the sensitivity to new physics in the case of large detector uncertainties as well as in the case of sizeable backgrounds. Moreover, ECO hunt and EPO hunt do not suffer from the traditional look-elsewhere effect. These findings are not limited to the case of proton beam-dump experiments, but will be relevant in any setting where backgrounds are rare and difficult to model and the optimal high-level observables are unknown. In particular, we expect the methods proposed here to be valuable also in searches for long-lived particles at the LHC.

To conclude, we emphasize that the approaches discussed here require the assumption of a specific model, because they exploit the predicted differential distributions to enhance sensitivity. An important avenue for future research would be to extend our analysis to a more model-independent approach, which remains agnostic regarding the production mechanism of the long-lived particles. In this case it may be useful to extend the EPO hunt to two-dimensional posteriors (for mass and lifetime) in order to extract the maximal amount of information from data.

\section*{Acknowledgements}

We are grateful to Stephen Jiggins and Jan Kieseler for valuable comments on earlier versions of the manuscript. We furthermore thank (in alphabetical order) Alexander Heidelbach, Ulrich Husemann, Markus Klute, Ulrich Nierste and Maksym Ovchynnikov for discussions. This work was supported by the Helmholtz-Gemeinschaft Deutscher Forschungszentren (HGF) through the Young
Investigators Group VH-NG-1303 and by the Deutsche Forschungsgemeinschaft (DFG) through Grant No. 396021762 -- TRR 257. 

\section*{Code and data}

The code used for the event generation, the network architecture and its training can be found at \url{https://github.com/amorandini/param_hunt}. An example Jupyter notebook which explains how to
derive the plots and results in this paper is also provided. Neural networks are built and trained with
Tensorflow \cite{tensorflow2015-whitepaper} and use Keras \cite{chollet2015keras} as backend. Hyperparameter scans are performed with KerasTuner \cite{omalley2019kerastuner}.

\appendix

\section{Network architecture, hyperparameters and validation}
\label{sec:appendix_architecture}

For the classifier that performs the posterior inference, we use the architecture summarised in \cref{tab:post_arch} and a training sample consisting of 250k events of different masses and lifetimes. We find that the lifetime posterior is typically rather flat and uninformative, and therefore focus on the marginalised posterior for the ALP mass. The architecture and hyperparameters of the classifier assessing the compatibility based on low-level observables is given in \cref{tab:cl_obs_arch}. Given that the posterior extraction from low-level observables and the compatibility assessment based on low-level observables are similar tasks, the architectures of the two classifiers are very similar. The only differences between the two are the sample sizes and the use of batch normalization. These differences do not significantly change the final performance, but make the training procedure more stable.

\setlength\extrarowheight{1.2mm}

\begin{table}[]
    \centering
    \caption{Architecture and hyperparameters of the classifier extracting posteriors. A training sample consisting of 250k events is used for training.}
    \label{tab:post_arch}
    \begin{tabular}{l c}
    \hline
    \hline
             Architecture &  \\
         \hline
         Number of hidden layers & 4\\
         Units per hidden layer & 64 \\
          Hidden layer activation function & prelu \\

         \hline \hline
          Hyperparameters &  \\
         \hline
         Batch size & 4096 \\
      Batch normalization & no \\

         Max number epochs & 100 \\
         Initial learning rate & $2.4\cdot 10^{-2}$ \\
         Decay rate & 0.95 every 500 steps \\
         Early stopping (loss) & $\delta < 5\cdot 10^{-3}$ for 20 epochs\\
         \hline\hline
    \end{tabular}
    \vspace{5mm}
    \centering
    \caption{Architecture and hyperparameters of the classifier applied to the low-level features. A training sample consisting of 500k events is used for training.}
    \label{tab:cl_obs_arch}
    \begin{tabular}{lc}
    \hline
    \hline
             Architecture &  \\
         \hline
         Number of hidden layers & 4\\
         Units per hidden layer & 64 \\
          Hidden layer activation function & prelu \\
         \hline \hline
          Hyperparameters &  \\
         \hline
         Batch size & 4096 \\
         Batch normalization & yes \\
         Max number epochs & 100 \\
         Initial learning rate & $2.4\cdot 10^{-3}$ \\
         Decay rate & 0.95 every 500 steps \\
         Early stopping (accuracy) & $\delta < 5\cdot 10^{-3}$ for 20 epochs\\ 
         \hline\hline
    \end{tabular}
    \vspace{5mm}
    \caption{Architecture and hyperparameters of the classifier applied to the event posteriors. The same training sample of the classifier applied to observable is used, but the posteriors have been extracted by a trained classifier. Posteriors are stacked to form a $(200,\,2)$ array which is later processed by 1D convolutional layers.}
    \label{tab:cl_post_arch}
    \begin{tabular}{lc}
    \hline\hline
             Architecture &  \\
         \hline
        Convolution 1D filters & [64, 64, 128]\\
        Convolution activation function & relu \\
        After convolutional layer & Max pooling \\
         Number of hidden layers & 4\\
         Units per hidden layer & 64 \\
         Hidden layer activation function & prelu \\
         \hline \hline
          Hyperparameters &  \\
         \hline
         Batch size & 4096 \\
          Batch normalization & no \\
         Max number epochs & 100 \\
         Initial learning rate & $5\cdot 10^{-4}$ \\
         Decay rate & 0.95 every 500 steps \\
         Early stopping (accuracy) & $\delta < 5\cdot 10^{-3}$ for 20 epochs\\             \hline\hline
    \end{tabular}

\end{table}

The architecture used to evaluate the compatibility between events based on their posteriors is given in \cref{tab:cl_post_arch}. In this case, the network input is very different and the architecture has been adapted accordingly. In particular, the input has higher dimensionality, as the posteriors are evaluated on a grid of 200 points in the ALP mass. We regularise the posterior probability by taking the natural logarithm, requiring values to be larger than $-4$ and normalising the result to lie between 0 and 1. The posteriors for the two events under consideration are then stacked to form a $(200,\,2)$ array. This kind of pre-processing is inspired by image recognition. The posteriors are then passed through a series of 1D convolutions until they reach dimension $(128,\,2)$ and finally a max pooling layer reducing them to $(128,\,1)$ dimensions. Now the resulting array is evaluated by a series of dense hidden layers until a compatibility score is returned. 

All hyperparameters have been selected after a series of hyperparameter scans. 
All the training curves show fast convergence and little to no overfitting. In particular for the classifier applied to event posteriors, the performance is very stable against changes in hyperparameters, because the task is comparably easy.

A general difficulty when using classifiers to infer posterior probabilities for model parameters is how to assess the network performance if the true posterior is unknown. One possibility to measure performance is to study the coverage, i.e.\ to determine how often the true parameters are contained in a credible interval of given probability \cite{hermans2021trust}. While we do not have access to the true posterior, we have access to the model parameters used to generate the events. We can use our estimated marginal posteriors to derive the minimal credible region that contains the ALP mass. This will constitute our empirical coverage that needs to be compared with the expected coverage.

\begin{figure}
    \centering
    \includegraphics[width=0.5\textwidth]{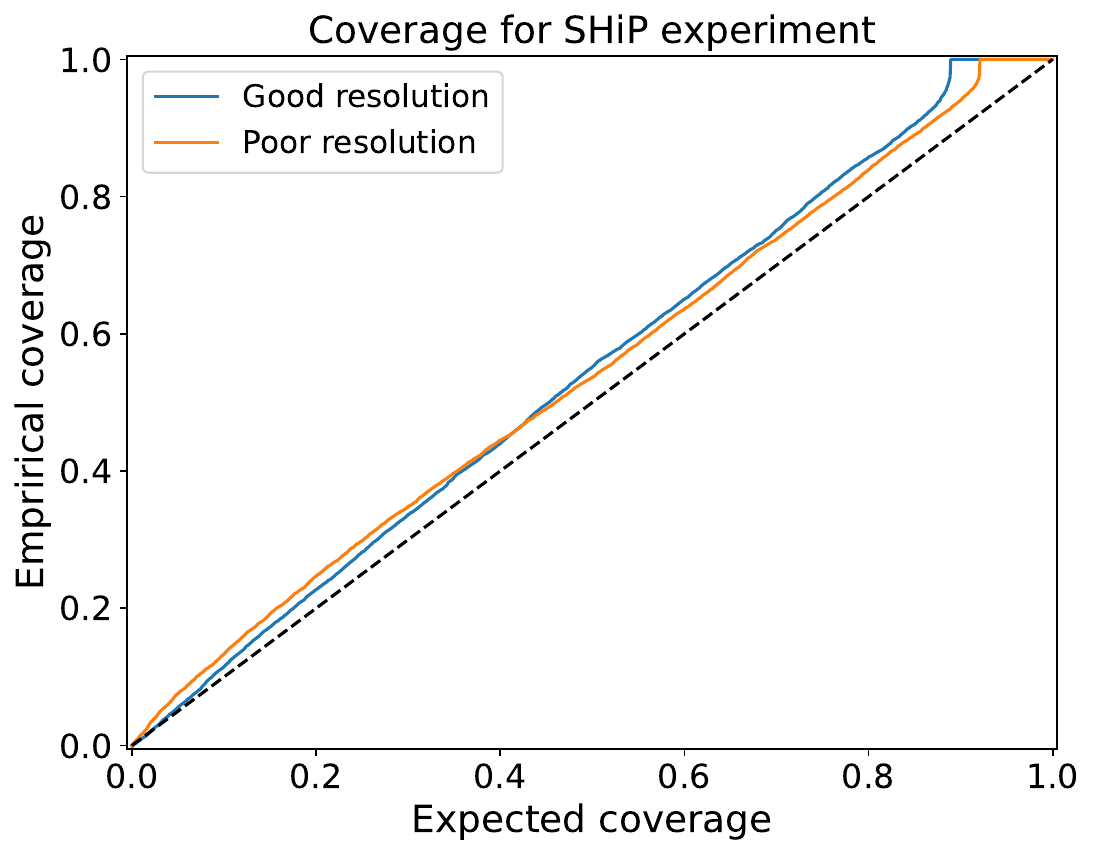}
    \caption{Coverage plots for the posteriors of our classifiers in the case of good and poor resolution. A small underconfidence is present. The plateau at large coverage is given by the fact that the true mass value falls within the bin with highest posterior.}
    \label{fig:coverage}
\end{figure}

In \cref{fig:coverage} we can see that our posterior estimates are slightly conservative in the sense that the uncertainties are slightly overestimated. Fig.~\ref{fig:coverage} can however not tell us whether the inferred posterior is optimal, i.e.\ as narrow as possible given the available information. To answer this question, we have considered an experiment with close to perfect energy resolution, such that the invariant mass of the diphoton pair correlates very tightly with the ALP mass. We find that in this case the posterior width agrees with the resolution of the diphoton invariant mass, confirming that our network performs close to optimally.  For the remainder of this work we will assume that this is also the case for detectors with poor energy resolution. In any case, for the purpose of assessing event compatibility, it is sufficient that the posterior estimates are not over-confident. 

\section{Threshold selection}\label{sec:appendix_bkgrej}

Throughout this study we have fixed the threshold of our compatibility algorithm to 0.68. This specific value was chosen to maximise the average sensitivity $\langle Z\rangle_s$ for all benchmark scenarios and across the whole range of signal strengths and background normalisation. 
In this appendix we will discuss how stable our performances are against the choice of threshold. We will show how the sensitivity $\langle Z\rangle_s$ varies for several physical scenarios when using different thresholds. In the second part of this appendix we will consider a slightly different task that requires a different performance measure. We will show that we obtain a similar threshold dependence of the performance.

\subsection{Optimizing for $\langle Z\rangle_s$ sensitivity}

\begin{figure*}
    \centering
    \includegraphics[width=0.8\textwidth]{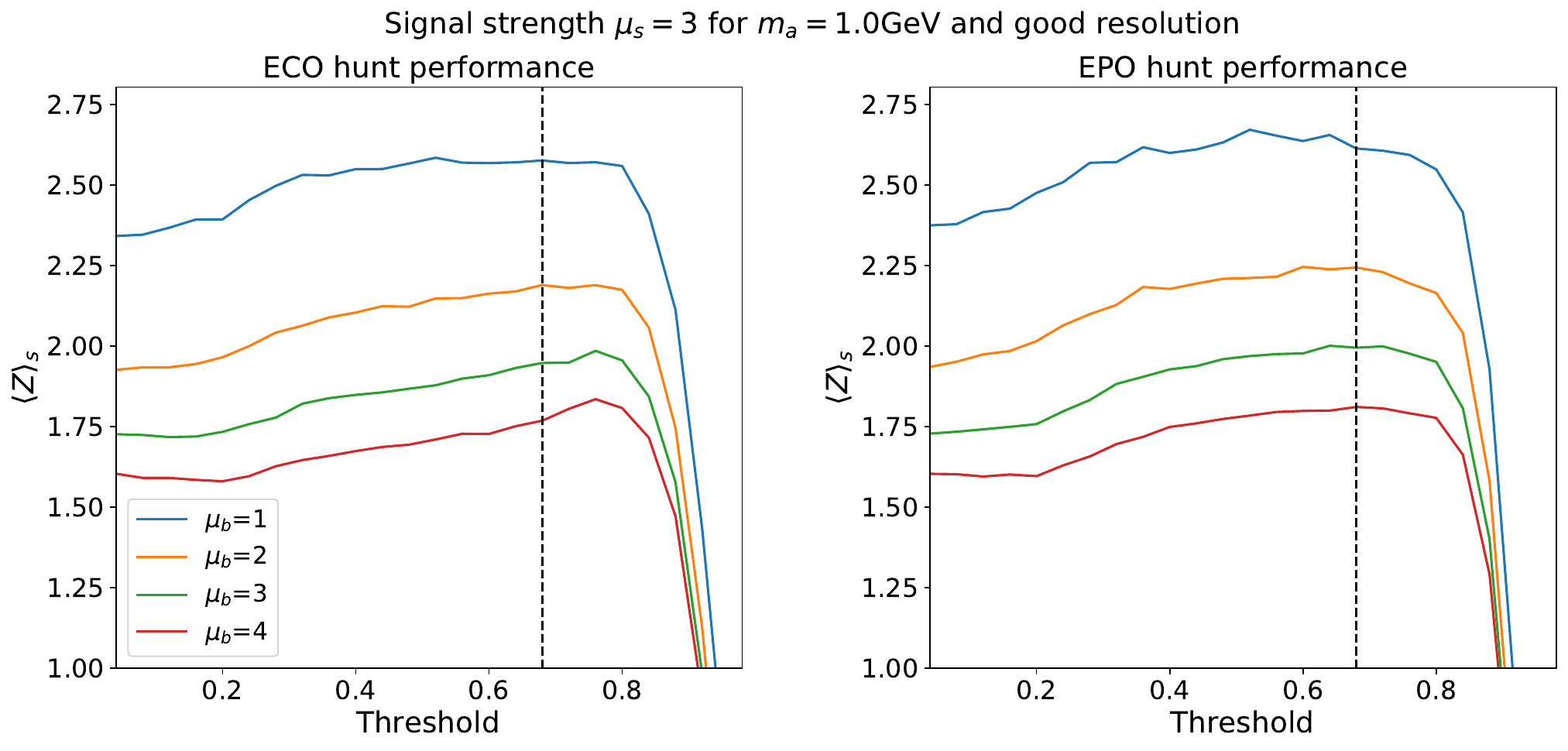}
        \includegraphics[width=0.8\textwidth]{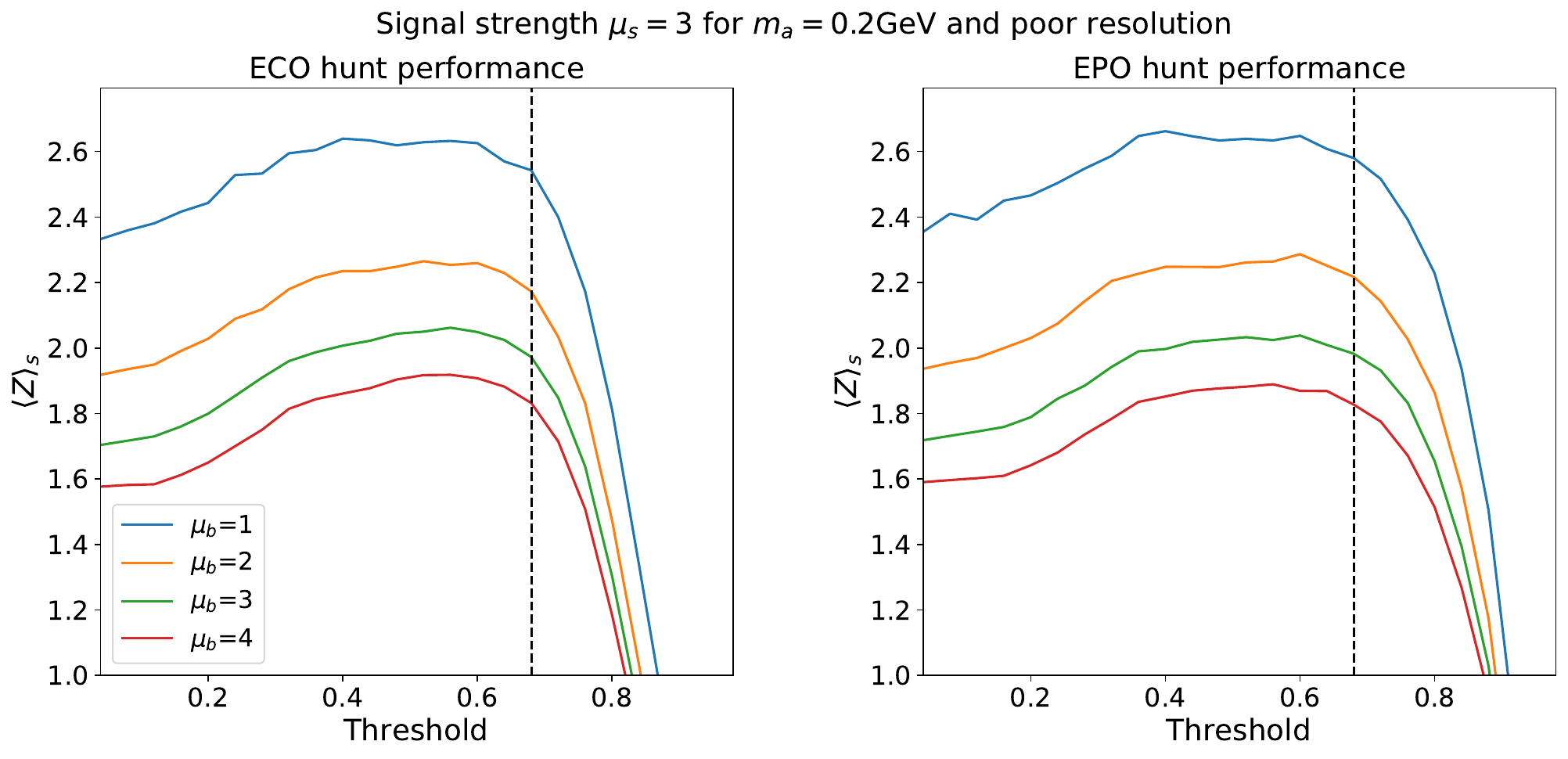}
    \caption{Value of the sensitivity for fixed signal strength and varying background normalization. The dashed vertical line indicates our optimal threshold choice.}\label{fig:z_thr}
\end{figure*}

We start by discussing the performance measure $\langle Z\rangle_s$ introduced in \cref{sec:performance_sensitivity}. In the main text we show the performances for fixed threshold over the ranges of $\mu_b$ and $\mu_s$ of interest. Here we provide the complementary information of how the performance varies for different thresholds for fixed values of $\mu_b$ and $\mu_s$. Our findings are summarised in \cref{fig:z_thr}.

Regardless of the physical scenario, we see similar trends in the performance dependence. The optimal threshold is always in the range $0.6-0.8$. Going to lower thresholds leads to a smoothly decreasing performance, while overshooting the optimal value leads to a steep decrease in performance. The optimal threshold also depends on the physical scenario in the expected way: large ALP masses seen by a detector with good resolution require a larger threshold than small masses seen by a detector with poor resolution. This is because large masses lead to more peaked posteriors, which are easier to distinguish from incompatible events, hence we can use a stricter compatibility requirement in our algorithm.

Overall, a threshold of 0.68 is a good compromise. While it is not the best choice for each scenario considered, it performs almost optimally for all of them. This choice of threshold is therefore appropriate for a model-independent exploratory study. It is of course possible to further optimise the threshold choice when more information about the physical scenario is available.

\subsection{Further background rejection performances}

In the main text we consider a binary classification of pairs of events into either compatible (i.e.\ signal events) or incompatible (i.e.\ background events). Here we consider a more difficult task, namely to perform classification between samples of $n_\text{obs} > 2$ events and determine whether or not background events are present in the sample. Specifically, we want to compare the case where $n_\text{sig}$ signal events and $n_\text{bkg}$ are observed to the case where no signal events and $n_\text{bkg}+n_\text{sig}$ background events are observed. For this task the signal strength and background normalization do not play any role in the sense that the total amount of observed events cannot be used to distinguish the background only and background plus signal scenarios.

We are interested in the distribution of the TS defined in section~\ref{sec:performance_sensitivity}, i.e.\ the number of events identified as signal by the ECO and EPO hunt. Let us call $p^{n_\text{obs}}_n(l)$ the distribution of the TS when there are $n_\text{obs}$ events in a sample of which $n$ are signal events. We are then interested in quantifying the difference between the distributions $p^{n_\text{obs}}_n(l)$ and $p^{n_\text{obs}}_0(l)$, i.e.\ how good we are at distinguishing a sample of $n_\text{obs}$ events that contains $n$ signal events from one that does not contain any. Similarly to before we can then define
\begin{equation}
    \langle L_{p,0} \rangle^{n_\text{obs}}_n\equiv \sum_{l=1}^{n_\text{obs}} \left(p^{n_\text{obs}}_n (l) \sum_{m\geq l}^{n_\text{obs}} L_{p,0}^{n_\text{obs}} (m)\right) \; ,
\end{equation}
where $L_{p,0}$ is the value of $L_p$ for $n = 0$. 
In contrast to the performance measure $\langle Z \rangle_s$, this quantity measures how well we discard background events, rather than quantifying the expected sensitivity. Differently from the AUC considered in \cref{sec:bkg_remove}, the performance measure is non-binary, as it depends on the total number of events.

\begin{figure*}
    \centering
    \includegraphics[width=0.8\textwidth]{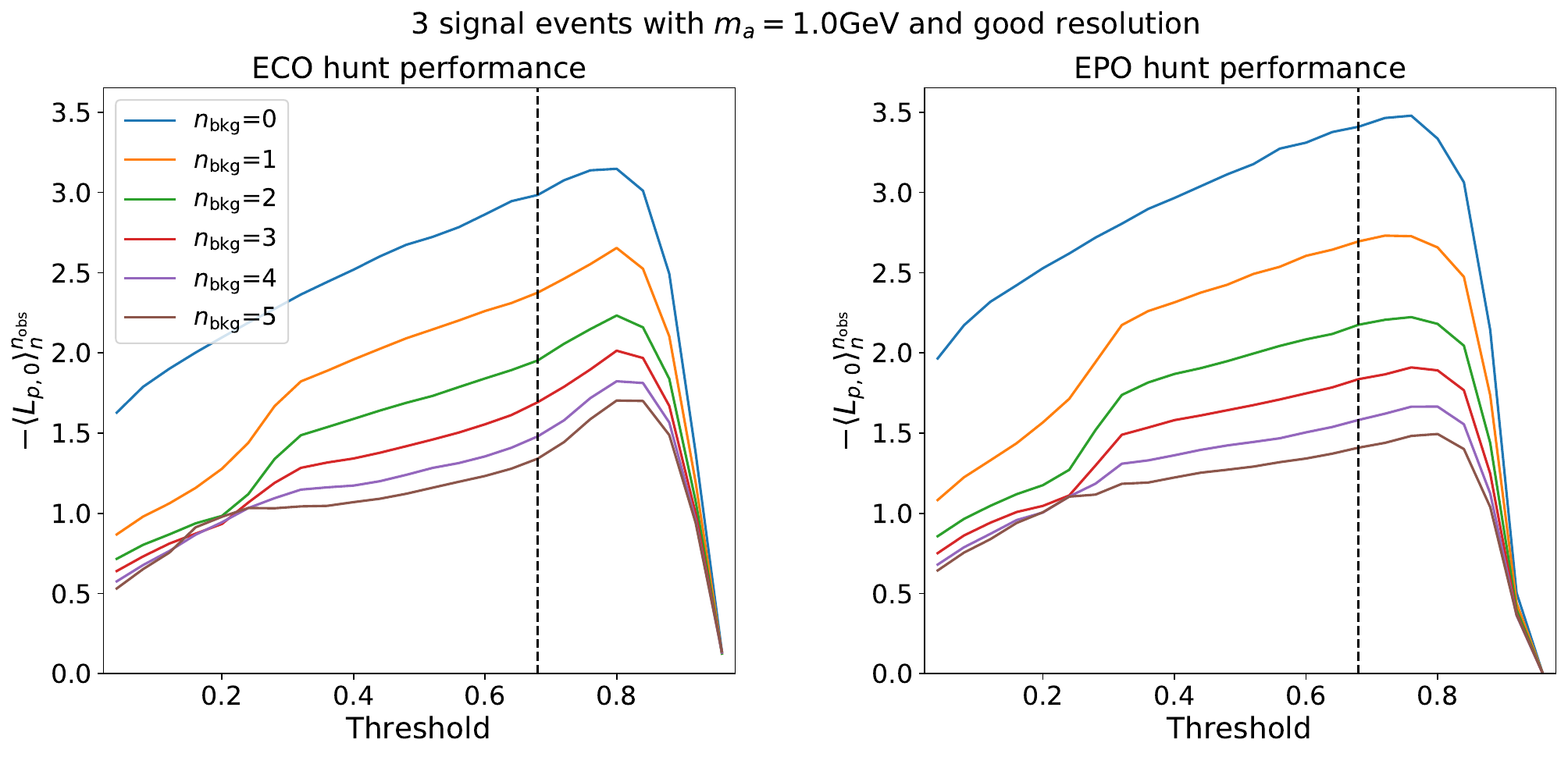}
        \includegraphics[width=0.8\textwidth]{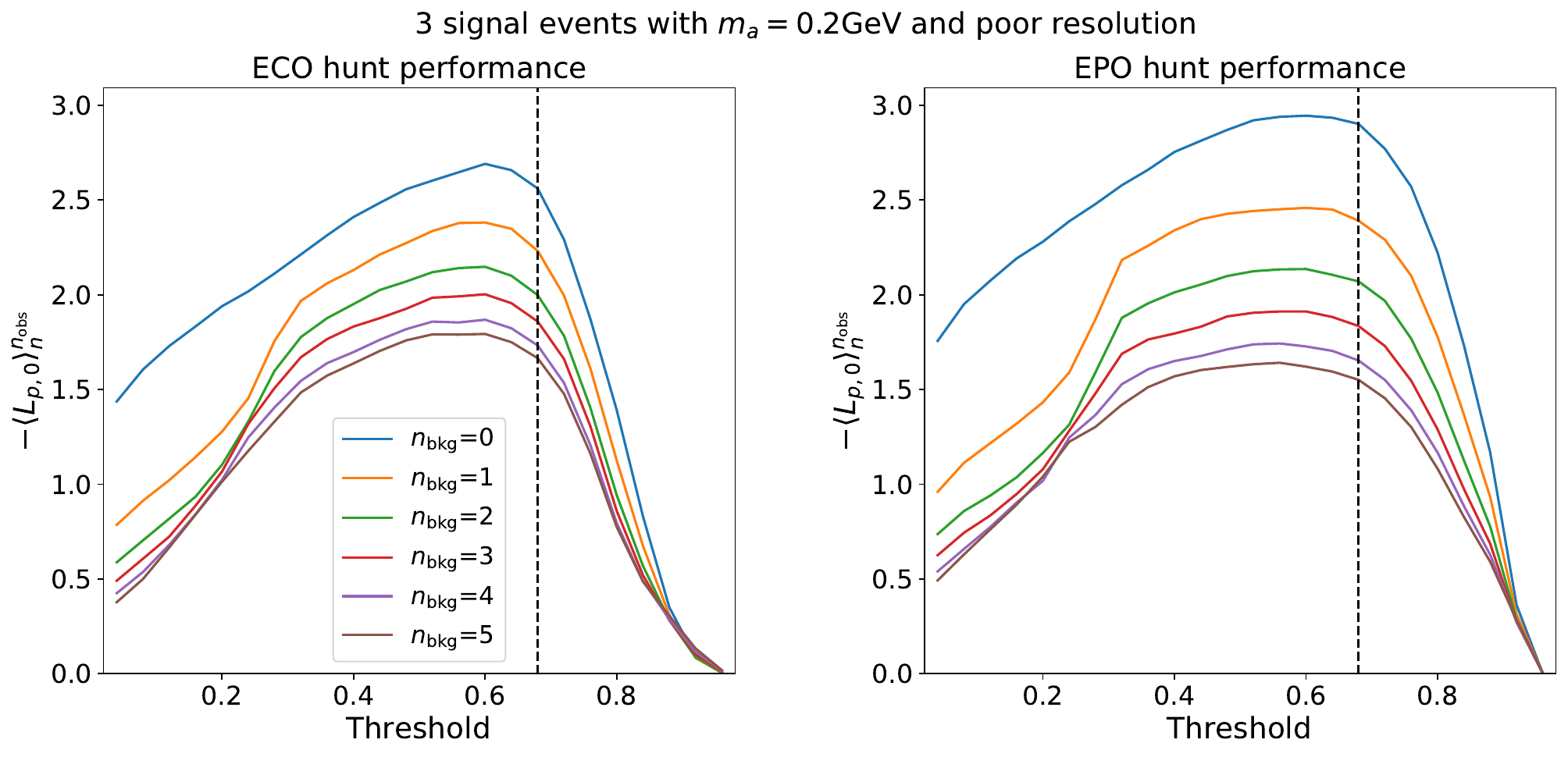}
    \caption{Mean value of $L_{p,0}$ for different cases of signal mass and detector resolution. }\label{fig:logp_thr}
\end{figure*}

We show this new performance measure for varying contamination of background events and for different threshold values in \cref{fig:logp_thr}.
As before, the optimal value of the threshold will depend on the position and spread of the signal. The dependence on the threshold choice is stronger than in the previous case, but it exhibits exactly the same trends. Taking a threshold which is too large leads to a steep decrease in performance, while taking lower thresholds leads to roughly linearly decreasing performances. 

We find that the threshold value of 0.68 is a reasonable choice also for this task, even though it has been selected to maximise $\langle Z\rangle_s$. 
This is a non-trivial result, because the task of identifying a signal excess and discarding background events are not the same, even though they are of course closely related. The fact that the same threshold works reasonably well for both the tasks highlights the robustness of our approach.

\interlinepenalty=10000

\bibliographystyle{bibstyle}
\bibliography{bibliography}

\end{document}